\documentclass[aps,reprint,nofootinbib,twocolumn,superscriptaddress,longbibliography]{revtex4-1}
\usepackage{subfigure}
\usepackage{amsmath, amssymb,graphicx,color,bm,epstopdf}
\usepackage[dvipsnames]{xcolor}
\usepackage{bbold}
\usepackage{braket}
\usepackage{physics}
\usepackage{comment}
\usepackage{natbib}
\usepackage{slashed}

\newcommand{\be}{\begin{equation}}
\newcommand{\e}{\end{equation}}
\newcommand{\beml}{\begin{subequations}}
\newcommand{\eml}{\end{subequations}}
\newcommand{\beq}{\begin{eqnarray}}
\newcommand{\eq}{\end{eqnarray}}
\newcommand{\ba}{\begin{array}}
\newcommand{\ea}{\end{array}}
\newcommand{\bpm}{\begin{pmatrix}}
\newcommand{\epm}{\end{pmatrix}}
\newcommand{\bc}{\begin{cases}}
\newcommand{\ec}{\end{cases}}

\usepackage{placeins}
\usepackage{float}
\usepackage{upgreek}

\usepackage{todonotes}

\usepackage{tabularx}
\usepackage{tikz}

\usetikzlibrary{calc,intersections,through,hobby}
\usepackage{cleveref}
\usetikzlibrary{decorations.pathmorphing}
\tikzset{snake it/.style={decorate, decoration=snake}}

\usepackage{epsfig}

\DeclareMathAlphabet{\oldcal}{OMS}{cmsy}{m}{n}

\newcommand{\tripleship}[3]{\langle#1\text{\textbar}#2\text{\textbar}#3\rangle}

\definecolor{amendments}{rgb}{0.0, 0.0, 0.7}

\newcommand{\compconj}[1]{%
  \overline{#1}%
}

\begin{document}

\title{Quantum electrodynamical density functional theory for generalized Dicke model}

\author{A.~Kudlis}
\email{andrewkudlis@gmail.com}
\affiliation{Abrikosov Center for Theoretical Physics, MIPT, Dolgoprudnyi, Moscow Region 141701, Russia}
\author{D.~Novokreschenov}
\affiliation{Faculty of Physics, ITMO University, St. Petersburg 197101, Russia} 
\author{I.~Iorsh}
\email{i.iorsh@metalab.ifmo.ru}
\affiliation{Abrikosov Center for Theoretical Physics, MIPT, Dolgoprudnyi, Moscow Region 141701, Russia}
\affiliation{Department of Physics, Engineering Physics and Astronomy, Queen’s University, Kingston, Ontario K7L 3N6, Canada }
\author{I.~V.~Tokatly}\affiliation{Nano-Bio Spectroscopy Group and European Theoretical Spectroscopy Facility (ETSF), Departamento de Polímeros y Materiales
Avanzados: Física, Química y Tecnología, Universidad del País Vasco, Avenida Tolosa 72, E-20018 San Sebastián, Spain}
\affiliation{IKERBASQUE, Basque Foundation for Science, 48009 Bilbao, Spain}
\affiliation{Donostia International Physics Center (DIPC), E-20018 Donostia-San Sebastián, Spain}
\affiliation{Faculty of Physics, ITMO University, St. Petersburg 197101, Russia}%

\date{\today}

\begin{abstract}
We formulate and analyze in detail the ground state quantum electrodynamical density functional theory (QEDFT) for a generalized Dicke model describing a collection of $N$ tight-binding dimers minimally coupled to a cavity photon mode. This model is aimed at capturing essential physics of molecules in quantum cavities in polaritonic chemistry, or quantum emitters embedded in mesoscopic resonators of the circuit QED, and, because of its simplicity, is expected to provide important insights regarding the general QEDFT. We adopt the adiabatic connection formalism and the diagrammatic many-body theory to regularly derive a sequence of explicit approximations for the exchange-correlation (xc) energy in the ground state QEDFT, and to compare their performance with the results of exact numerical diagonalization. Specifically, we analyze the earlier proposed one-photon optimized effective potential (OEP) scheme, its direct second order extensions, and a non-perturbative xc functional based on the photon random phase approximation (RPA). Our results demonstrate the excellent performance of RPA-QEDFT in the ultrastrong coupling regime, and for any number $N$ of Dicke molecules in the cavity. We study in detail the scaling of xc energy with $N$, and emphasize the importance for the ground state QEDFT of collective effects in the interaction of molecules with cavity photons. Finally, we discuss implications of our results for realistic systems.  
\end{abstract}

\maketitle
\section{Introduction}

The last decade witnessed a substantial  progress in photonic technologies, allowing to routinely fabricate micro- and nanocavities characterized both by large quality factors and extremely small mode volumes, and therefore, deeply subwavelength field localization. This stimulated the research in cavity quantum electrodynamics (QED), where a novel,  \textit{utrastrong},  regime of light matter coupling~\cite{Kockum2019} has been actively explored. In the ultrastrong coupling regime, the characteristic energy of light-matter interaction becomes comparable to the cavity photon frequency: this results in the increased role of the vacuum electromagnetic fluctuations, which may modify the ground state of the matter placed inside the cavity. Specifically, utrastrong coupling was predicted to induce various cavity mediated phase transitions ~\cite{thomas2019exploring,curtis2019cavity,sentef2018cavity,schlawin2019cavity,li2020manipulating,ashida2020quantum,PhysRevLett.125.257604, wang2019cavity} and to substantially modify the chemical reactions~\cite{martinez2018can}. The latter phenomenon led to the emergence of the new interdisciplinary research direction, polaritonic or QED chemistry~\cite{Ebbesen2016,ebbesen2016hybrid,GarciaVidal2021} exploring the influence of the electromagnetic vacuum fluctuations on the chemical properties of the cavity-embedded matter. 

For the quantitative predictions of the cavity mediated chemical properties of real materials, one should resort to the ab-initio modelling. Specifically, the generalization of density function theory (DFT)~\cite{HohKohn1964,DreizlerGross1990,RunGro1984,TDDFTbyUllrich} accounting for the fluctuating quantum electromagnetic, termed quantum electrodynamical DFT (QEDFT)~\cite{Tokatly2013PRL,Ruggenthaler2014PRA,ruggenthaler2017groundstate}, has been developed over the last decade, in parallel with many other many-body techniques including self-consistent (Hartree-Fock) field methods~\cite{Vendrell2018,RivFliNar2019,Haugland2021}, coupled cluster theory~\cite{Haugland2020,Mordovina2020,Haugland2021,DePrince2021,Liebenthal2022}, diagrammatic methods~\cite{TreMil2015,MelMar2016,Tokatly2018PRB}, configuration integration approaches~\cite{Ahrens2021,McTague2022}, and very recently the diffusion quantum Monte Carlo technique \cite{Weight2023arxiv-2}.

Among different many-body techniques, QEDFT looks especially promising, especially for complex systems, because by analogy to the standard electronic DFT, it is expected to provide a good balance between the accuracy and numerical efficiency. It also demonstrates a sufficient flexibility to address the ground state problems \cite{Flick2018} and the problems of dynamics, including analysis of excitations in the linear response \cite{Flick2019,Yang2021,Yang2022} as well as the modeling of cavity mediated chemical reactivity \cite{Schafer2022b}. It can be also adopted to multimode and lossy cavities to describe natural line width of excitations, radiation losses and related dissipative dynamics \cite{Wang2021,Kudlis2022PRB,Schafer2022}. Recently, QEDFT has been extended to the case of cavities treated within the so-called macroscopic QED parametrizing realistic optical setups \cite{Svendsen2023arxiv}.

As in any DFT-type approach, a practical application of QEDFT relies on approximations for exchange correlation (xc) potential that encodes all complicated many-body effects. Unfortunately, for the moment little is known about properties of xc functionals responsible photon-mediated electronic correlations in the cavity QED. There are only a few attempts to regularly address the problem of approximations in QEDFT \cite{Pellegrini2015PRL,Flick2018,Flick2022,Schafer2021pnas}. In practice, only a perturbative first order optimized effective potential OEP of Ref.~\cite{Pellegrini2015PRL} has been formulated in a practical form and tested for dynamical problems in a number of model systems \cite{Flick2015,Flick2017a,Kudlis2022PRB}, and in the ground state version for realistic molecules \cite{Flick2018}. Recently, a promising local density version of the one-photon perturbative xc functional has been proposed for the ground state QEDFT \cite{Flick2022}. 

One obvious problem of the currently available approximations, based in the first order OEP, is a missing collective behaviour in the case if several chemically independent molecules are present in the cavity and interact with the same photon mode. For the ground state this implies that in these approximations the electron density in a given molecule is not influenced by the other molecules in the cavity, which does not look corect in the cavity QED context. However how important this collectivity is in reality, in particular for the ground state QEDFT, and which improvemens of xc functionals are needed to describe the corresponding physics is curently unknown. This work is aimed at addressing these questions and providing at least partial answers, using a simple model system as an example.

In this paper, we focus on the ground state QEDFT applied to a special case of generalized Dicke model that is one of the most known model of quantum optics used in many different contexts in the cavity and circuit QED \cite{PhysRevA.94.033850,pilar2020thermodynamics,PhysRevA.97.043820, 10.21468/SciPostPhys.9.5.066,lamata2017digital, shapiro2020universal,PhysRevA.100.022513, akbari2023generalized}. The specific version of the Dicke model we adopt in this work can be understood physically as describing $N$ two-site tight-binding molecules (dimers) interacting with a cavity photon mode via a minimal gauge invariant coupling. This system is a cartoon of a typical theoretical setup of polaritonic chemistry, in which a set of chemically independent molecules interacting with cavity photons is described by the Pauli-Fitz Hamiltonian within the dipole approximation. It therefore perfectly suits to our purpose of deriving, comparing and testing different explicit approximate xc functionals, and analyzing the relevance of quantum collectivity in the ground state QEDFT. As we will see, one of the beautiful simplifications we have in the Dicke model is that regular diagrammatic many-body methods of constructing DFT approximations produce explicit and frequently analytic density dependence, while in general they always generate OEP orbital functionals as for example in Refs.~\cite{Casida1995,Barth2005,Tokatly2018PRB}.

The paper is organized as follows. To make the paper self-contained, in Sec.~II we present a simple and compact quantization of Maxwell equations in the dipole approximation, which directly leads to dipole-gauge Hamiltonian used as a starting point in most first-principal approaches to QED chemistry. In Sec.~III the adiabatic connection formalism is adopted to derive the exact representation for xc energy in the ground state QEDFT. Here we also introduce the specific approximations analyzed in this work, the one- and two-photon perturbative OEP, and a non-pertirbative photon RPA functional. Section~IV presents main results of this work. Here we apply the general formalism introduced in Sec.~IV, to generalized Dicke model. By starting with the simplest case of one dimer, we then go to the general multi-dimer situation, analyze different approximation and compare their performance with the results of exact numerical diagonalization of the Dicke Hamiltonioan. The main results and general conclusions of this work are sumarized in Sec.~V.  

\section{Quantization of electromagnetic field: Hamiltonian for cavity QED}

Our aim is to describe a system of non-relativistic electrons strongly coupled to quantum electromagnetic modes of a microcavity, that is the main object of the cavity QED. In a typical setup the size of the electronic subsystem is much smaller that the wavelength of relevant photon modes, which justifies the use of the dipole approximation for the electron-photon interaction. The corresponding effective Hamiltonian is usually derived from the Pauli-Fierz Hamiltonian by performing a transformation to the dipole Power–Zienau–Woolley (PZW) gauge \cite{PowZie1959,Woolley1971,BabLou1983}. To reveal the physics behind the PZW Hamiltonian it is instructive to assume the dipole coupling at the level of Maxwell equations for the fields strengths and quantizing them directly \cite{AbeKhoTok2018EPJB}.  

Let us start from the Maxwell equations for the transverse part of the electromagnetic filed
\begin{eqnarray}
\nabla\times{\bf E}_{\perp} & = & -\frac{1}{c}\partial_{t}{\bf B},\label{Maxwell-EB-1}\\
\nabla\times{\bf B} & = & \frac{1}{c}\partial_{t}{\bf E}_{\perp}+\frac{4\pi}{c}{\bf j}_{\perp},\label{Maxwell-EB-2}
\end{eqnarray}
where ${\bf E}_{\perp}({\bf r},t)$ is the transverse component of the electric field
with $\nabla\cdot{\bf E}_{\perp}=0,$ and ${\bf j}_{\perp}({\bf r},t)$
is the transverse part of electron current that enters as a source
of the radiation field. The dipole approximation corresponds to approximating the current by a delta-function with a strength equal to the time derivative of the total dipole moment of the electronic system,
\begin{equation}
{\bf j}({\bf r},t)= \partial_{t}{\bf P}({\bf r},t) = e\dot{\bf R}(t)\delta({\bf r}-{\bf r}_{0}),\label{current}
\end{equation}
where ${\bf P}({\bf r},t)$ is the polarization, ${\bf R}=\sum_{j=1}^{N}{\bf r}_{j}$ is the center-of-mass position of the electrons, and it assumed that coordinates ${\bf r}_{j}$ of all $N$ electrons are bounded to
a region around some point ${\bf r}_{0}$ inside the cavity, which is much smaller than the cavity size and thus much smaller than the characteristic wavelength $\lambda$ of the field.

The transverse current coupled to the cavity modes via the Maxwell equations is given
by the transverse projection of the polarization vector, 
\begin{align}
    &{\bf P}_{\perp}({\bf r},t)=e{\bf R}(t)\delta^{\perp}({\bf r}-{\bf r}_{0})\nonumber\\&=\frac{e}{4\pi}\nabla\times\left(\nabla\times\frac{{\bf R}(t)}{|{\bf r}-{\bf r}_{0}|}\right)\label{P-transverse}
\end{align}

In a quantum theory the Maxwell equations (\ref{Maxwell-EB-1}), (\ref{Maxwell-EB-2})
should correspond to Heisenberg equations for field operators. The Hamiltonian structure of these equations is revealed by introducing a new electric variable -- the displacement vector, 
\begin{equation}
{\bf D_{\perp}}={\bf E}_{\perp}+4\pi{\bf P}_{\perp},\label{D}
\end{equation}
and rewriting the Maxwell equations(\ref{Maxwell-EB-1}), (\ref{Maxwell-EB-2}) as follows 
\begin{eqnarray}
\partial_{t}{\bf B} & = & -c\nabla\times({\bf D}_{\perp}-4\pi{\bf P}_{\perp}),\label{Maxwell-DB-1}\\
\partial_{t}{\bf D}_{\perp} & = & c\nabla\times{\bf B}.\label{Maxwell-DB-2}
\end{eqnarray}
One can easily check that by considering the standard energy of the transverse electromagnetic field
\begin{align}
&\hat{H}_{\text{e-m}}=\frac{1}{8\pi}\int d{\bf r}\left[\hat{\bf E}_{\perp}^{2}+\hat{\bf B}^{2}\right]\nonumber\\&=\frac{1}{8\pi}\int d{\bf r}\left[(\hat{\bf D}_{\perp}-4\pi\hat{\bf P}_{\perp})^{2}+\hat{\bf B}^{2}\right],\label{He-m}
\end{align}
supplemented with the following commutation relations for the components of the magnetic field and the electric displacement operators\footnote{Throughout this paper we use a system of units with $\hbar=1$.}
\begin{equation}
[\hat{B}^{i}({\bf r}),\hat{D}_{\perp}^{j}({\bf r}')]=-i4\pi c\varepsilon^{ijk}\partial_{k}\delta({\bf r}-{\bf r}'),\label{BD-commut}
\end{equation}
we recover the Maxwell equations \ref{Maxwell-EB-2}-\eqref{Maxwell-DB-2} from the canonical Heisenberg equations
\begin{eqnarray}
\partial_{t}\hat{\bf B} & = & i[\hat{H}_{\text{e-m}},\hat{\bf D}_{\perp}],\label{Maxwell-H-1}\\
\partial_{t}\hat{\bf D}_{\perp} & = & i[\hat{H}_{\text{e-m}},\hat{\bf B}].\label{Maxwell-H-2}
\end{eqnarray}
This analysis clearly shoes that the proper conjugated
Hamiltonian variables for the electromagnetic filed are the magnetic
filed ${\bf B}$ and the electric displacement ${\bf D}$ \cite{Cohen-Tannoudji-book}. 

The next step is to introduce the cavity mode functions as a set of normalized transverse eigenfunctions
${\bf f}_{\alpha}({\bf r})$ of the wave equation inside a metallic
cavity $\Omega$
\begin{eqnarray*}
c^{2}\nabla^{2}{\bf f}_{\alpha}({\bf r}) & = & \omega_{\alpha}^{2}{\bf f}_{\alpha}({\bf r}),\quad{\bf r}\in\Omega\\
({\bf n\times{\bf f}_{\alpha}})\vert_{\partial\Omega} & = & 0,
\end{eqnarray*}
where ${\bf n}$ is a unit vector normal to the cavity surface $\partial\Omega$. Such defined mode functions ${\bf f}_{\alpha}({\bf r})$ are proportional to the electric field in the $\alpha$-mode, but because the normalization to unity they are have a dimension of the square root of inverse volume. The amplitude of the mode function can thus be regarded as the inverse square root of the "mode volume". 

The basis set of the mode functions can be used to represent all transverse functions in the Hamiltonian (\ref{He-m})
\begin{eqnarray}
\hat{\bf D}_{\perp}({\bf r}) & = & \sum_{\alpha}\hat{d}_{\alpha}{\bf f}_{\alpha}({\bf r}),\label{D-expansion}\\
\hat{\bf B}({\bf r}) & = & \sum_{\alpha}\hat{b}_{\alpha}\frac{c}{\omega_{\alpha}}\nabla\times{\bf f}_{\alpha}({\bf r}),\label{B-expansion}\\
\hat{\bf P}_{\perp}({\bf r}) & = & e\sum_{\alpha}\left({\bf f}_{\alpha}({\bf r}_{0})\cdot\hat{\bf R}\right){\bf f}_{\alpha}({\bf r}).\label{P-expansion}
\end{eqnarray}
Here the expansion coefficients $d_{\alpha}$ and $b_{\alpha}$ are,
respectively, the quantum amplitudes of the electric displacement
and the magnetic field in the $\alpha$-mode. By substituting the
expansions of Eqs.~\eqref{D-expansion}-\eqref{P-expansion} into Eqs.~(\ref{He-m}) and (\ref{BD-commut}) we obtain the following Hamiltonian
\begin{equation}
\hat{H}_{\text{e-m}}=\frac{1}{8\pi}\sum_{\alpha}\left[\left(\hat{d}_{\alpha}-4\pi e{\bf f}_{\alpha}({\bf r}_{0})\cdot\hat{\bf R}\right)^{2}+\hat{b}_{\alpha}^{2}\right],\label{H-bd}
\end{equation}
and the commutations relations for the field amplitudes
\begin{equation}
[\hat{b}_{\alpha},\hat{d}_{\beta}]=-i4\pi\omega_{\alpha}\delta_{\alpha\beta}.\label{bd-commut}
\end{equation}

Finally we rescale the electric displacement and the magnetic field amplitudes
\begin{equation}
\hat{d}_{\alpha}=\sqrt{4\pi}\omega_{\alpha}\hat{q}_{\alpha},\quad \hat{b}_{\alpha}=\sqrt{4\pi}\hat{p}_{\alpha},\label{pq-def}
\end{equation}
so that the new variables $\hat{q}_{\alpha}$ and $\hat{p}_{\alpha}$ satisfy
the standard coordinate-momentum commutation relations $[\hat{p}_{\alpha},\hat{q}_{\beta}]=-i\delta_{\alpha\beta}$,
while the Hamiltonian (\ref{H-bd}) reduces to that for a set of shifted
harmonic oscillators
\begin{equation}
\hat{H}_{\text{e-m}}=\frac{1}{2}\sum_{\alpha}\left[\hat{p}_{\alpha}^{2}+\omega_{\alpha}^{2}\left(\hat{q}_{\alpha}-\frac{\bm{\lambda}_{\alpha}}{\omega_{\alpha}}\hat{\bf R}\right)^{2}\right],\label{Hem-PZW}
\end{equation}
where the ``coupling constant'' $\bm{\lambda}_{\alpha}$ is related
to the electric field of the $\alpha$-mode at the location of the
electron system
\begin{equation}
\bm{\lambda}_{\alpha}=\sqrt{4\pi}e{\bf f}_{\alpha}({\bf r}_{0}).\label{lambda}
\end{equation}
It is worth noting that proportionality of $\bm{\lambda}_{\alpha}$ to the mode function assumes the coupling strength inversely proportional to the square root of the mode volume.

The Hamiltonian of electromagnetic filed in Eq.~(\ref{Hem-PZW}) corresponds to the description of
electron-photon coupling in the PZW gauge \cite{PowZie1959,Woolley1971,BabLou1983,Cohen-Tannoudji-book}. The total Hamiltonian for the combined system of electrons and the field is a sum of $H_{\text{e-m}}$
(\ref{Hem-PZW}) and the standard Hamiltonian of a nonrelativistic
many-electron system
\begin{equation}
\hat{H}=\hat{T}+\hat{V}_{ext}+\hat{W}_{C}+\hat{H}_{\text{e-m}}.\label{H-fin}
\end{equation}
Here $\hat{T}$ corresponds to the kinetic energy of the electrons, $\hat{V}_{ext}=\sum_{j=1}^Nv_{\textup{ext}}(\mathbf{r}_j)$ the energy of interaction with an external potintial, and $\hat{W}_{C}=\frac{1}{2}\sum_{i\ne j}\frac{e^2}{|\mathbf{r_i}-\mathbf{r_j}|}$ is the energy of the Coulomb electron-electron interaction. This Hamiltonian is commonly used as the starting point in first-principles approaches to the cavity
QED, in particular in QEDFT which we discuss in the next section. 

\section{Energetics of electron-photon system in QEDFT}

In the discussion below we focus of the ground state QEDFT assuming the validity of the dipole approximation, which is the most common situation in cavity QED. In this case, described by the Hamiltonian of Eq.~\eqref{H-fin}, the standard Hohenberg-Kohn mapping theorem \cite{HohKohn1964} can be adopted to the cavity QED setup practically without modifications \cite{ruggenthaler2017groundstate} (the general ground state QEDFT beyond dipole approximation is discussed in Ref.~\cite{ruggenthaler2017groundstate}, see also Ref.~\cite{Penz2023arxiv}).   

\subsection{The ground state xc energy: Adiabatic connection formulation}

The Hohenberg-Kohn theorem adopted to the cavity QED Hamiltonian Eq.~\eqref{H-fin} implies that the ground state energy $E_0=\langle\Psi_0|\hat{H}|\Psi_0\rangle$ of the interacting electron-photon system can be represented as follows,
\begin{equation}
    \label{F-def}
    E_0 = F[n] + \int v_{\textup{ext}}(\mathbf{r})n(\mathbf{r}),
\end{equation}
where $F[n]$ is a universal Hohenberg-Kohn functional. The central object of any ground-state DFT, the xc energy functional $E_{xc}[n]$, is defined via the Kohn-Sham (KS) construction with respect to the KS system \cite{DreizlerGross1990}. In the QEDFT context the KS system corresponds to a system of free photons and non-interacting KS particles in the presence of an effective KS potential $v_s$ that is adjusted to reproduce the exact interacting electron density $n(\mathbf{r})$ in the non-interacting KS system. The xc energy is defined by the following alternative representation of the interacting ground state energy,
\begin{equation}
    \label{Exc-def}
    E_0 = T_s + \frac{1}{2}\sum_{\alpha}\omega_{\alpha} + \int v_{\textup{ext}}(\mathbf{r})n(\mathbf{r}) + E_H + E_{xc},
\end{equation}
where $T_s$ is the kinetic energy of KS particles and as usual the Hartree energy $E_H$ related to the Coulomb interaction is separated explicitly. Importantly, the mean-field contribution related to the electron-photon interaction does not appear in the ground state energy because the expectation value of the cavity transverse electric field $\hat{\bm{e}}_{\alpha} = \mathbf{f}_{\alpha}(\mathbf{r}_0)(\hat{d}_\alpha - 4\pi e\mathbf{f}_{\alpha}(\mathbf{r}_0)\cdot\hat{\mathbf{R}})$ in the equilibrium state vanishes identically, 
\begin{equation}\label{zero-e}
\langle\hat{\bm{e}}_{\alpha}\rangle =\frac{1}{e}\langle\bm{\lambda}_\alpha(\omega_{\alpha}\hat{q}_{\alpha}-\bm{\lambda}_{\alpha}\hat{{\bf R}}) \rangle = 0
\end{equation} 
It is worth noticing that this identity fixes the expectation value of the photon coordinate $\langle\hat{q}_\alpha\rangle$ (the average electric displacement in the $\alpha$-mode) by relating it to the average dipole moment of the electrons $e\langle\hat{\mathbf{R}}\rangle=e\int\mathbf{r}n(\mathbf{r})d\mathbf{r}$ that is a simple explicit functional of the density.

To obtain a formally exact representation of the xc energy in terms of linear response functions we use the standard adiabatic connection fluctuation-dissipation (ACFD) formalism \cite{Gunnarsson1976,Langreth1977,Dobson-in-TDDFT-2012,TDDFTbyUllrich}. Recently ACFD machinery has been already applied in the QEDFT context \cite{Flick2022}. Here we use a slightly different formulation by keeping density fixed along the adiabatic path \cite{Gunnarsson1976,Langreth1977}, and including, on equal footing, both electron-photon coupling and the direct electron-electron interaction.  

Let us introduce an adiabatic parameter $0<\gamma<1$, and rescale the electron-electron interaction and the electron-photon coupling  such that $\gamma=0$
and $\gamma=1$ correspond to completely decoupled and fully interacting systems. In addition we enforce the electron density $n(\mathbf{r})$ to be the same at any value of $\gamma$ by properly adjusting an external one-particle potential $\hat{V}_\gamma = \sum_{j=1}^Nv_\gamma(\mathbf{r}_j)$. Apparently at $\gamma =1$ the potential $\hat{V}_{\gamma=1}=\hat{V}_{ext}$ coincides with the physical external potential in the fully intercating system, $\hat{V}_{\gamma=0}=\hat{V}_{s}$ is equal to the KS potential and the system at $\gamma=0$ is actually the KS system of free photons and non-interacting KS particles. Formally, the electron-photon system along this adiabatic path in described by the following Hamiltonian,
\begin{align}\nonumber
\hat{H}_{\gamma}=\hat{T} &+ \hat{V}_\gamma + \gamma\hat{W}_C  \\
 &+ \frac{1}{2}\sum_{\alpha}\left[\hat{p}_{\alpha}^{2}+\omega_{\alpha}^{2}\left(\hat{q}_{\alpha}-\gamma\frac{\bm{\lambda}_{\alpha}}{\omega_{\alpha}}{\bf \hat{R}}\right)^{2}\right]\label{H-adiabatic}
\end{align}

The ground state energy $E_0=E_0^{\gamma=1}$ of fully interacting system can then be computed using the Hellmann-Feynman theorem,
\begin{align}
E_{0} &= E_{0}^{\gamma=0} + \int_{0}^{1}d\gamma\langle\Psi_{0}^{\gamma}|\frac{\partial\hat{H}_{\gamma}}{\partial \gamma}|\Psi_{0}^{\gamma}\rangle = T_s + V_{ext}\nonumber\\
&+ \int_{0}^{1}d\gamma\langle\Psi_{0}^{\gamma}|\hat{W}_C - \frac{e}{\gamma}\sum_{\alpha}{\hat{\bf R}}\cdot\hat{\bm{e}}_{\alpha}^{\gamma}|\Psi_{0}^{\gamma}\rangle\label{E0-ACFD}
\end{align}
where $|\Psi_{0}^{\gamma}\rangle$ is the ground state wave function, and $\hat{\bm{e}}_{\alpha}^{\gamma}=\gamma\bm{\lambda}_{\alpha}(\omega_{\alpha}\hat{q}_{\alpha}-\gamma\bm{\lambda}_{\alpha}\hat{{\bf R}})$ is the electric field operator for the system described by the Hamiltonian of Eq.~\eqref{H-adiabatic}.

By comparing Eqs.~\eqref{Exc-def} and \eqref{E0-ACFD} we can identify the xc energy and observe that it naturally splits into two contributions attributed, respectivelly, to the direct electron-electron interaction and to the electron photon coupling\footnote{This is similar to the exact representation of xc force in the time-dependent version of QEDFT also known as QED-TDDFT \cite{Tokatly2013PRL}.},
\begin{equation}
    \label{Exc-total}
    E_{xc} = E_{xc}^{\rm el} + E_{xc}^{\rm ph}
\end{equation}
The electronic part comes $E_{xc}^{\rm el}$ from $\hat{W}_C$ term in Eq.~\eqref{E0-ACFD}. It is given by the standard ADFD formula and can be conveniently expressed in terms of the density-density response function $\chi_n^\gamma(\mathbf{r},\mathbf{r}',\omega)=\langle\langle\delta\hat{n}(\mathbf{r});\delta\hat{n}(\mathbf{r}')\rangle\rangle_\omega$ (see, for example Refs.~\cite{Dobson-in-TDDFT-2012,TDDFTbyUllrich}),
\begin{align}    \nonumber
 E_{xc}^{\rm el} &= \int_{0}^{1}d\gamma\langle\Psi_{0}^{\gamma}|\hat{W}_C |\Psi_{0}^{\gamma}\rangle - E_H \\
 \nonumber
 &= -\frac{1}{2}\int_{0}^{1}d\gamma\int d\mathbf{r}d\mathbf{r'}
 \frac{e^2}{|\mathbf{r} - \mathbf{r}'|}\\ \label{Exc-el}
 &\times \left[\int\frac{d\omega}{\pi}\chi_n^\gamma(\mathbf{r},\mathbf{r}',i\omega) + \delta(\mathbf{r} - \mathbf{r}')n(\mathbf{r}) \right].
\end{align}
The electron-photon part $E_{xc}^{\rm el}$ of xc energy originates from the last term in Eq.~\eqref{E0-ACFD} and is determined by the static correlation function of the electronic dipole moment and the cavity electric field, $\langle\Psi_{0}^{\gamma}|{\bf \hat{R}}\cdot\hat{\bm{e}}_{\alpha}^\gamma|\Psi_{0}^{\gamma}\rangle$. The fluctuation-dissipation theorem then relates the static correlation function to the corresponding Kubo response function $\chi_{{\bf R,}\bm{e}\alpha}^\gamma=\langle\langle{\bf \hat{R}}\cdot\hat{\bm{e}}_{\alpha}^\gamma\rangle\rangle$
as follows,
\begin{align}\nonumber
&\langle\Psi_{0}^{\gamma}|{\bf \hat{R}}\cdot\hat{\bm{e}}_{\alpha}^\gamma|\Psi_{0}^{\gamma}\rangle = \sum_{n\ne0}\langle\Psi_{0}^{\gamma}|{\bf \hat{R}}|\Psi_{n}^{\gamma}\rangle\langle\Psi_{n}^{\gamma}| \hat{\bm{e}}_{\alpha}^\gamma|\Psi_{0}^{\gamma}\rangle \\
&= -\int_{0}^{\infty}\frac{d\omega}{\pi}{\rm Im}\chi_{{\bf R},\bm{e}\alpha}^\gamma(\omega)
= -\int_{0}^{\infty}\frac{d\omega}{\pi}\chi_{{\bf R},\bm{e}\alpha}^\gamma(i\omega)
\label{GS-correlator1}
\end{align}
This identity together with Eq.~\eqref{E0-ACFD} leads to the final ACFD representation of the photon contribution to the xc energy,
\begin{align}
E_{xc}^{\rm ph} = e\int_{0}^{1}\frac{d\gamma}{\gamma}\int_{0}^{\infty}\frac{d\omega}{\pi} 
\sum_{\alpha}\chi_{{\bf R},\bm{e}\alpha}^{\gamma}(i\omega)
\label{Exc-ph}
\end{align}
Therefore the cavity induced correction to the xc energy is determined by the off-diagonal response function $\chi_{{\bf R},\bm{e}\alpha}(\omega)$ that describes the cavity electric field generated by an external classical field applied to the electrons.  By the Onsager reciprocity, this response function also determines the polarization of the electronic system induced by external currents (external dipole antennas) driving the cavity degrees of freedom.

\subsection{Relation to the electron polarizability and the photon propagator}

Let us apply to the electrons a classical time-dependent uniform electric
field ${\bf E}_{ext}(\omega)$ which, by polarizing the electronic system,
induces the cavity electric field,
\[
{\bf E}_{ext}(\omega)\to\delta{\bf R}(\omega)\to\bm{e}_{\alpha}(\omega)
\]
The response function relating the net (summed over all modes) cavity electric field  $\bm{e}(\omega)$ to the external field ${\bf E}_{ext}(\omega)$
is the quantity we need to compute the photon contribution to the xc energy Eq.~(\ref{Exc-ph}),
\begin{align}
e^{i}(\omega) &=\sum_{\alpha}e_{\alpha}^{i}(\omega) = -e\,\sum_{\alpha}\chi_{e_{\alpha}^{i},R^{j}}(\omega)E_{ext}^{j}(\omega)\nonumber\\
&\equiv -e\,\chi_{e^{i},R^{j}}(\omega)E_{ext}^{j}(\omega)\label{e-Eext-response-def}
\end{align}

The required off-diagonal response function $\chi_{e^{i},R^{j}}(\omega)$ can be conveniently expressed in terms of the polarizability of the electron subsystem and the photon proparator. 

We define a 1-photon irreducible electronic polarizability
$\Pi(\omega)$ as a polarization response of the electrons to the total
electric field, that is, the sum of the external and the cavity fields, 
\begin{equation}
\delta R^{i}(\omega)=e\Pi^{ij}(\omega)[E_{ext}^{j}(\omega)+e^{j}(\omega)]
\label{proper-Pi-def}
\end{equation}
The Hamiltonian of Eq.~\eqref{H-fin} generates the following equation of motion for the expectation value of the photon coordinate $q_{\alpha}$,
\[
\ddot{q}_{\alpha}=-\omega_{\alpha}(\omega_{\alpha}q_{\text{\ensuremath{\alpha}}}-\lambda_{\alpha}^{k}R^{k}).
\]
By using the definition of the cavity field, $e_{\alpha}^{i}=\frac{1}{e}\lambda_{\alpha}^{i}(\omega_{\alpha}q_{\text{\ensuremath{\alpha}}}-\lambda_{\alpha}^{k}R^{k})$, this equation
can be transformed into the equation for $\bm{e}_{\alpha}(t)$,
\begin{equation}
\ddot{e}_{\alpha}^{i}+\omega_{\alpha}^{2}e_{\alpha}^{i}=-\frac{1}{e} \lambda_{\alpha}^{i}\lambda_{\alpha}^{k}\ddot{R}^{k},
\label{mode-Maxwell}
\end{equation}
which is nothing, but the $\alpha$-mode projected Maxwell equation for the cavity electric field driven by the transverse part of the electron current. On the right hand side of Eq.~\eqref{mode-Maxwell}, $R^{k}(t)$ can be replaced by $\delta R^{k}(t)$ because only the induced dynamical part of electronic polarization generates the cavity field. From here we find the Fourier components of the electric field in each mode,
\[
e_{\alpha}^{i}(\omega)=- \frac{1}{e}\frac{\lambda_{\alpha}^{i}\omega^{2}\lambda_{\alpha}^{k}}{\omega^{2}-\omega_{\alpha}^{2}}\delta R^{k}(\omega),
\]
and the net cavity field,
\begin{equation}
e^{i}(\omega)=\sum_{\alpha}e_{\alpha}^{i}(\omega) =
\frac{1}{e}D^{ik}(\omega)\delta R^{k}(\omega)
\label{e-R-response}
\end{equation}
Here we defined the function $D^{ik}(\omega)$ that is a bare propagator
of the cavity electric field evaluated at the location of the electronic
system,
\begin{equation}\label{e-propagator}
D^{ik}(\omega)=-\sum_{\alpha}\frac{\lambda_{\alpha}^{i}\omega^{2}\lambda_{\alpha}^{k}}{\omega^{2}-\omega_{\alpha}^{2}}
\end{equation}

Finally, by inserting Eq.(\ref{proper-Pi-def}) into Eq.(\ref{e-R-response})
we get the equation relating the cavity field to the external field,
\[
\left[\delta_{ij}-D^{ik}(\omega)\Pi^{kj}(\omega)\right]e^{j}(\omega)=D^{ik}(\omega)\Pi^{kj}(\omega)E_{ext}^{j}(\omega),
\]
which gives the desired response function,
\begin{equation}
e\,\chi_{{\bm e},{\bf R}}(\omega)= - \left[1-\hat{D}(\omega)\hat{\Pi}(\omega)\right]^{-1}\hat{D}(\omega)\hat{\Pi}(\omega)
\label{e-Eext-response-fin}
\end{equation}
This results can be represented as a product $\chi_{{\bm e},{\bm R}}(\omega)=-e\,\hat{\mathcal{D}}(\omega)\hat{\Pi}(\omega)$ of the fully dressed, physical propagator of the cavity field $\hat{\mathcal{D}}=(1-\hat{D}\hat{\Pi})^{-1}\hat{D}$, and the 1-photon irreducible electron polarizability $\hat{\Pi}$. Diagrammatically, the correlation function
$\chi_{{\bf R},\bm{e}\alpha}^{g}(i\omega)$ of Eq.~\eqref{e-Eext-response-fin} corresponds to the sum of all polarization diagrams describing the propagation and screening of the cavity electric $\bm{e}_{\alpha}$ induced by a time-dependent classical electric field applied to the electronic subsystem.

To compute the photon contribution to the xc energy  within the adiabatic connection scheme we rescale all $\lambda_{\alpha}$ with the adiabatic factor of $\gamma$, that is, $\lambda_{\alpha}\to \gamma\lambda_{\alpha}$, modify accordingly the response function Eq.~\eqref{e-Eext-response-fin}, and substitute the result into Eq.~\eqref{Exc-ph},
\begin{align}
&E_{xc}^{\rm ph}  =-\int_{0}^{\infty}\frac{d\omega}{\pi}\int_{0}^{1}\frac{d\gamma}{\gamma}\nonumber\\
&\times{\rm tr}\left\{ \left[1-\gamma^{2}\hat{D}(i\omega)\hat{\Pi}_\gamma(i\omega)\right]^{-1}\gamma^{2}\hat{D}(i\omega)\hat{\Pi}_\gamma(i\omega)\right\}
\label{Exc-ph-fin}
\end{align}
Here $\hat{\Pi}_\gamma(\omega)$ is the exact 1-photon irreducible polarizability of the electronic subsystem with the rescaled interaction. The polarizability defined via Eq.~\eqref{proper-Pi-def} can be expressed in terms of the can be expressed in terms of the density-density response function entering the electronic part Eq.~\eqref{Exc-el} of the xc energy,
$$
\Pi^{ij}_\gamma(\omega) = - \int r_1^i\chi_n^\gamma(\mathbf{r}_1,\mathbf{r}_2,\omega)r_2^j d\mathbf{r}_1d\mathbf{r}_2.
$$
Therefore Eq.~\eqref{Exc-ph-fin} provides the exact representation of $E_{xc}^{\rm ph}$ in terms of the same object -- the density response function, that determines the electronic contribution $E_{xc}^{\rm el}$ given by Eq.~\eqref{Exc-el}.   

\subsection{Approximations for xc energy: photon RPA}

The exact ACFD representation Eq.~\eqref{Exc-ph-fin} for the photonic part of xc energy is a convenient starting point for constructing approximate functionals. In general, a variety of approximations can be generated by using different perturbative forms of the electronic polarizability and/or a perturbative expansion of Eq.~\eqref{Exc-ph-fin} itself.  

In particular, by approximating the polarizability $\hat{\Pi}_\gamma$ with skeleton diagrams constructed from the electron KS Green function $G_s$, the bare photon propagators, and, if relevant, the direct Coulomb interaction, we can generate a sequence of OEP-type conserving approximations for QEDFT \cite{Tokatly2018PRB}. This construction is an extension of the diagrammatic OEP construction in the electronic DFT \cite{Barth2005,Casida1995}. The simplest approximation in this sequence is the 1-photon OEP proposed in Ref.~\cite{Pellegrini2015PRL}. It corresponds to keeping the lowest order in the photon propagator term in Eq.~\eqref{Exc-ph-fin}, that is $E_{xc}^{\rm 1ph}= \frac{1}{2}{\rm tr}[\hat{D}\hat{\Pi}_s]$, where $\hat{\Pi}_s=\hat{\Pi}_{\gamma=0}$ is the polarizability of the KS system. This approximation, as well as its more recent LDA version \cite{Flick2022} may have potentially a serious problem when applied to realistic setups in QED chemistry which typically deals with many well separated molecules in a cavity. In this situation, the irreducible polarizability scales linearly with the number $N$ of molecules and so does the 1-photon xc energy, $E_{xc}^{\rm 1ph}\sim N$. This apparent extensivity of the xc energy leads to the xc potential inside a given molecule, which does not depend on $N$ and thus on the presence of the other molecules in the cavity. However, the latter is not what one would expect physically because of the extreme long-range character and nonextensivity of the cavity induced interaction.

\begin{figure}[t!]
\centering
\begin{align*}
   \!\!E_{xc}^{\rm ph}=-\dfrac{1}{2} \, \vcenter{\hbox{\begin{tikzpicture}[thick, use Hobby shortcut, scale=0.38]
\draw (-1.8,0) .. (0.0,-1.8) .. (1.8,0);
\draw (-1.8,0) .. (0.0,1.8) .. (1.8,0);
\path [draw=black,snake it]
    (-1.8,0) -- (1.8,-0);
 \fill (-1.8,0) circle (4pt);
 \fill (1.8,0) circle (4pt);
 \end{tikzpicture}}}-\dfrac{1}{4} \,   \vcenter{\hbox{\begin{tikzpicture}[thick, use Hobby shortcut, scale=0.38]
\draw (-1.8,-1.7) .. (-2.6,0.0) .. (-1.8,1.7);
\draw (-1.8,-1.7) .. (-1.,0.0) .. (-1.8,1.7);
  \path [draw=black,snake it]
    (-1.8,-1.7) -- (1.8,-1.7);
      \path [draw=black,snake it]
    (-1.8,1.7) -- (1.8,1.7);
\draw (1.8,-1.7) .. (2.6,0.0) .. (1.8,1.7);
\draw (1.8,-1.7) .. (1.,0.0) .. (1.8,1.7);
 \fill (1.8,-1.7) circle (4pt);
 \fill (1.8,1.7) circle (4pt);
  \fill (-1.8,-1.7) circle (4pt);
 \fill (-1.8,1.7) circle (4pt);
 \end{tikzpicture}}}-\dfrac{1}{4} \, \vcenter{\hbox{\begin{tikzpicture}[thick, use Hobby shortcut, scale=0.38]
\draw (-1.8,0) .. (0.0,-1.8) .. (1.8,0);
\draw (-1.8,0) .. (0.0,1.8) .. (1.8,0);
\path [draw=black,snake it]
    (-1.27279,1.27279) -- (1.27279,-1.27279);
\path [draw=black,snake it]
    (-1.27279,-1.27279) -- (1.27279,1.27279);
 \fill (1.27279,1.27279) circle (4pt);
 \fill (-1.27279,1.27279) circle (4pt);
 \fill (1.27279,-1.27279) circle (4pt);
 \fill (-1.27279,-1.27279) circle (4pt);
 \end{tikzpicture}}}+\dots
\end{align*}
    \caption{The diagrammatic expansion for xc energy up to the second order in the photon propagator. The first two diagrams belong to RPA-series dominating for large $N$, while the last one is of exchange type.}
    \label{fig:basic_diag}
\end{figure}
The problem with scaling becomes obvious already at the second order of the perturbation theory, shown diagrammatically in Fig.~\ref{fig:basic_diag}. The last, second order exchange diagram contains one fermionic loop and is linear in $N$. In contrast, the second, RPA-type ring diagram involves two polarization loops, scales with $N^2$, and dominates provided $N$ is sufficiently large. From a similar argument, applied to an arbitrary order of the perturbation theory, shows that for large $N$, within each given order in the coupling constant, the RPA ring diagrams dominate because they contain maximal number of the electron polarization loops. Obviously, all these leading contributions strongly violate the artificial extensivity of the lowest OEP.  

Selection and summation of the dominant ring diagrams gives a new approximate xc energy functional, which we call the photon RPA. At the level of the ACFD expression Eq.~\eqref{Exc-ph-fin}, this summation corresponds to approximating the exact irreducible polarizability with that of the KS system $\hat{\Pi}_{\gamma}\approx\hat{\Pi}_s$, and performing the coupling constant integration. The result of this procedure reads,       
\begin{equation}
    \label{Exc-ph-RPA}
 E_{xc}^{\rm ph} \approx E_{xc}^{\rm RPA} = \int_{0}^{\infty}\frac{d\omega}{2\pi}{\rm tr}\ln\left[1-\hat{D}(i\omega)\hat{\Pi}_s(i\omega)\right].
\end{equation}
This functional is the natural QED extension of the RPA correlation energy used in the electronic DFT \cite{Dobson2005,Furche2008}.

We note that the argumentation based on the $1/N$ small parameter has been recently applied to several problems of cavity QED \cite{Dmytruk2021,Dmytruk2022,Lenk2022,Kudlis2023arxiv,Vlasiuk2023PRB}. In the next section, we study its relevance for the ground state QEDFT using a paradigmatic example of the Dicke model. 

\section{Photon RPA xc functional for the Dicke model}
\label{sec:one_dim}

\subsection{Formulation of the model}

Here, we apply the formalism developed in the previous section to construct and test the accuracy of several xc functionals for a model system of $N$ tight-binding dimers (two-site molecules) interacting with one cavity mode. In the site basis, $i^{\textup{th}}$ dimer is represented by a 2$\times$2 Hamiltonian with the hopping kinetic energy proportional to the $\hat{\sigma}_x^i$ Pauli matrix. The matrix $\hat{\sigma}_z^i$ represents the dimer's dipole moment (charge disbalance between the sites) that is coupled to the cavity electric field and to the external classical potential. The Hamiltonian of this system reads,
\begin{multline}
\hat{H} = -\sum_{i=1}^NT\hat{\sigma}_x^i + 
\sum_{i=1}^N v_\text{ext}^i\hat{\sigma}_z^i  \\
 + \frac{1}{2}\!\left[\hat{p}^2\!+\omega_{\textup{ph}}^2\left(\hat{q} -\dfrac{\lambda}{\omega_{\textup{ph}}}\sum_{i=1}^N\hat{\sigma}_z^i\right)^2\right],
  \label{H-Dicke-def}  
\end{multline}
where $\omega_\text{ph}$ is the frequency of the photon mode, and $T$ is the amplitude of the hopping between the sites of the dimers, which for simplicity are assumed to be identical. In the QEDFT context, the first and the second terms in Eq.~\eqref{H-Dicke-def} correspond, respectively, to the kinetic energy $\hat{T}$ and the external potential energy $\hat{V}_\text{ext}$ in the generic Hamiltonian of Eq.~\eqref{H-fin}. The last term in  Eq.~\eqref{H-Dicke-def} is apparently the $\hat{H}_\text{e-m}$ in Eq.~\eqref{H-fin} with the total dipole moment $\hat{\mathbf{R}}=\sum_{i=1}^N\hat{\sigma}_z^i$ equal to the sum of the dipole moments for all dimers. The expectation value of the density disbalance in the dimer $n_i=\langle\hat{\sigma}_z^i\rangle$ plays a role of basic density variable in QEDFT for this model system. This local density variable $n_i$ is conjugated to the external potential $v_\text{ext}^i$ in Eq.~\eqref{H-Dicke-def}. As usual in the lattice formulation of DFT, all density functionals of the general QEDFT now become functions of $N$ variables, $F[n(\mathbf{r})]\mapsto F(n_1,\dots n_N)$.

The model described by the Hamiltonian Eq.~\eqref{H-Dicke-def} is a cartoon of a typical polaritonic chemistry setup -- a multi-emitter cavity QED system. Formally, it is a special, gauge invariant version of the generalized Dicke model \cite{PhysRevA.94.033850,pilar2020thermodynamics,PhysRevA.97.043820, 10.21468/SciPostPhys.9.5.066,lamata2017digital, shapiro2020universal,PhysRevA.100.022513, akbari2023generalized}, which describes a set of two-level systems coupled to a bosonic mode via a minimal electromagnetic coupling. In fact, Eq.~\eqref{H-Dicke-def} is gauge equivalent to the Hamiltonian, where a coupling to the photon mode is introduced via Peierls substitution of the vector potential into the hopping phase. This gauge invariance prevents the unphysical superradince of the standard Dicke model \cite{Andolina2019}. 

\subsection{One dimer: QEDFT for the quantum Rabi model}

We start with the simple case of one dimer interacting with a single cavity mode,  that is, $N=1$. In this case, the generalized Dicke model defined by Eq.~\eqref{H-Dicke-def} reduces to the quantum Rabi model,
 \begin{equation}\label{eqn:gen_ham}
\hat{H}=\!-T\hat{\sigma}_x+v_{\textup{ext}}\hat{\sigma}_{z}+\frac{1}{2}\!\left[\hat{p}^2\!+\omega_{\textup{ph}}^2\left(\hat{q}-\dfrac{\lambda}{\omega_{\textup{ph}}}\hat{\sigma}_z\right)^2\right],
\end{equation}
which is probably one of the most famous models in quantum optics. In particular, it has been used in the QEDFT context to assess the validity of the 1-photon OEP~\cite{Pellegrini2015PRL}. In this section, we go beyond this simplest approximation and analyze xc functionals based on the second order of the selfconsitent perturbation theory, and on the full photon RPA of Eq.~\eqref{Exc-ph-RPA}.

It is natural to rewrite the photon canonical variables in the second quantization formalism as follows,
\begin{eqnarray} \label{photon-operators}
&&q=\dfrac{1}{\sqrt{2\omega_{\textup{ph}}}}\left(\hat{a}^{}+\hat{a}^{\dagger}\right),\ \ 
p=-i\sqrt{\dfrac{\omega_{\textup{ph}}}{2}}\left(\hat{a}^{\dagger}-\hat{a}^{}\right).\quad
\end{eqnarray}
In terms the operators $\hat{a}^{\dagger}$ and $\hat{a}$ the Hamiltonian reads
\begin{multline}
        \hat{H}=-T\hat{\sigma}_x+\left[\sqrt{\frac{\omega_{\textup{ph}}}{2}}\lambda(\hat{a}+\hat{a}^{\dagger})+v_{\textup{ext}}\right]\hat{\sigma}_{z}\\+\omega_{\textup{ph}}(\hat{a}^{\dagger}\hat{a}^{}+1/2)+\frac{\lambda^2}{2}.\label{hamiltonian_rabi}
\end{multline}
The eigenvalues of this model can be found by appropriately truncating the photon Fock subspace. As in this work we are concerned with the ground state QEDFT, only  the ground state of the system is required.

The KS system for the Rabi model correspond to one KS particle on a dimer in the presence of the KS potential $v_s=v_{\textup{xc}}+v_{\textup{ext}}$. The ground $|\phi_g\rangle$ and excited $|\phi_e\rangle$ KS orbitals, and the corresponding energy eigenvalues $\varepsilon_{e,g}$ are determined from the following KS equation,
\begin{align}\label{eqn:KS_auxiliary}
    \Big[-T\hat{\sigma}_x+v_s\hat{\sigma_{z}}\Big]|\phi_{e,g}\rangle=\varepsilon_{e,g}|\phi_{e,g}\rangle.
\end{align}
The solution of this simple two-level problem is readily found and reads, $\langle\phi_g|=(\compconj{v},\compconj{u})$ with $\varepsilon_g=-W$, and $\langle\phi_e|=(\compconj{u},-\compconj{v})$ with $\varepsilon_e=W$, where $W=\sqrt{v_s^2+T^2}$, and $\compconj{u},\compconj{v}=\sqrt{(1\pm v_s/W)/2}$. It is also convenient to introduce the resonant excitation frequency of the noninteracting KS system: $\Omega_s=2W$.

The above solution of the KS problem is represented as a function of the potential $v_s$ that, by the DFT mapping theorem, is a unique functional of the density. A remarkable feature of the two-level system is the functional relation between the potential $v_s$ and the basic density variable (the density disbalance of dimer nodes) $n=\tripleship{\phi_g}{\hat{\sigma}_z}{\phi_g}$ can be found explicitly, 
\begin{align}
    n &= -\frac{v_s}{W} = -\frac{v_s}{\sqrt{v_s^2+T^2}}, \label{n(v)} \\
   v_s &=-\dfrac{ n T}{\sqrt{1 - n^2}}. \label{v(n)}
\end{align}
The function $W$ which determines the KS eigenvalues in terms of $n$ reads as $W(n,T)= T/\sqrt{1 - n^2}$. Using $v_s(n)$ of Eq.~\eqref{v(n)} we can express the KS orbitals as the functions of $n$, and eventually find the explicit form of the KS kinetic energy functional $T_s(n)$ entering the ground state energy Eq.~\eqref{Exc-def}, 
\begin{align}\label{Ts-Rabi}
\!\!\!T_s(n)\!=\!-T\tripleship{\phi_g(n)}{\hat{\sigma}_x}{\phi_g(n)}\! =\!-T\sqrt{1-n^2},
\end{align}
where we used $\tripleship{\phi_g}{\hat{\sigma}_x}{\phi_g} = 2 u v = T/W=\sqrt{1-n^2}$. The final expression for the ground state energy of one dimer coupled to the cavity mode takes the form 
\begin{align}\label{E0-Rabi}
E_0=-T\sqrt{1-n^2}+\dfrac{\omega_{\textup{ph}}}{2}+v_{\textup{ext}}n + E_{xc}^{\rm ph}(n).
\end{align}
Here the xc energy $E_{xc}^{\rm ph}(n)$ is also naturally expressed as a function of the density if we adopt the diagrammatic representation of the ADFT formalism as described above in Sec.~III~C. The reason is the known $n$-dependence of the KS Green functions $G_s$ appearing in the diagrams for the xc energy, 
\begin{equation}
    \label{G_s(n)}
    G_s[n](\omega) = \left[\omega + T\hat{\sigma}_x - v_s(n)\hat{\sigma}_z\right]^{-1},
\end{equation}
where $v_s(n)$ is given by Eq.~\eqref{v(n)}.

Below, we analyze and compare several specific approximations based on the diagrammatic expansion for photon part of the xc energy $E_{xc}^{\rm ph}$. 

Let us start with the simplest finite order perturbative approximations. Figure~\ref{fig:basic_diag} shows the lowest perturbative contributions to $E_{xc}^{\rm ph}$, up to the second order in the photon propagator. The analytical expressions for the diagrams on Fig.~\ref{fig:basic_diag} denoted as $D_1$, $D_2$, and $D_3$, respectively, are given in Appendix~A, Eqs.~(\ref{D1}) - (\ref{D3}).  The first order diagram $D_1$ corresponds to the lowest order photon OEP functional, $E_{xc}^{\rm 1ph}(n)=D_1(n)$, proposed in Ref.~\cite{Pellegrini2015PRL}. The sum of all three diagrams on Fig.~\ref{fig:basic_diag} yields the complete second order extension of OEP, $E_{xc}^{\rm 2ph}(n)=D_1 + D_2 + D_3$, while the first two diagrams correspond to the photon RPA functional truncated at the second order, $E_{xc}^{\rm 2RPA}=D_1+D_2$.

The highest level xc functional we consider in this work is the photon RPA of Eq.~\eqref{Exc-ph-RPA}, which corresponds to the summation of all RPA-type ring diagrams up to infinite order. In the present case of the Rabi model the corresponding xc energy can be found explicitly as follows,
\begin{multline} \label{Exc-RPA-Rabi}
E_{xc}^{\rm RPA}=\int\limits_{0}^{\infty}\dfrac{d\omega}{2\pi}\textup{ln}\left[1-\lambda^2 D_{\textup{ph}}(i\omega)\Pi_s(i\omega)\right]
\\ \quad\quad =\dfrac{1}{2}\left[(\Omega_+ + \Omega_-) - (\omega_{\textup{ph}} + \Omega_s)\right],
\end{multline}
where, for convenience, we extracted the coupling constant dependence from the propagator of Eq.~\eqref{e-propagator}, and defined the bare photon propagator $D_\textup{ph}(\omega)$ as follows, 
\begin{align}
D_{\textup{ph}}(\omega)&=-\frac{\omega^2}{\omega^2-\omega_{\textup{ph}}^2}, \label{Dph}   
\end{align}
The KS polarizability $\Pi_s(\omega)$ of the dimer reads (see Appendix~A),
\begin{align}
\Pi_s(\omega)=-\dfrac{Z_s^{\textup{ex}}}{\omega^2-\Omega_s^2},
\end{align}
with the following oscillator strength of the KS exciton, $Z_s^{\textup{ex}}=4T^2/W=4T\sqrt{1-n^2}$. 

The result of the $\omega$-integration in Eq.~\eqref{Exc-RPA-Rabi} is expressed in terms of the frequencies, $\Omega_{+}$ and $\Omega_{-}$, of the upper and lower Rabi polaritons, 
\begin{multline}\label{eqn:polariton_requenc_one_dimer}
\Omega_{\pm}=\dfrac{1}{2}\left[\sqrt{(\omega_{\textup{ph}}+\Omega_0)^2+\lambda^2 Z_s^{\rm ex}}\right.\\ \pm \left.\sqrt{(\omega_{\textup{ph}}-\Omega_0)^2+\lambda^2 Z_s^{\textup{ex}}}\right].
\end{multline}
These frequencies correspond to zeroes of the expression $1-\lambda^2 D_{\textup{ph}}\Pi_s(\omega)$ and determine the energies of the physical bosonic excitations of the system in the presence of the electron-photon coupling. In particular, the polaritonic frequencies appear as poles in the physical dressed photon propagator,  
\begin{multline}\label{eqn:dressed_prop}
\mathcal{D}(\omega)=\dfrac{D_{\textup{ph}}(\omega)}{1-\lambda^2 D_{\textup{ph}}(\omega)\Pi(\omega)}=\\
=-\omega^2\left(\dfrac{Z^{\textup{ph}}_+}{\omega^2-\Omega_{+}^2}+\dfrac{Z^{\textup{ph}}_-}{\omega^2-\Omega_{-}^2}\right),
\end{multline}
where the strength $Z_{\pm}^{\textup{ph}}$ of the upper/lower polariton contribution to the physical photon is defined as,
\begin{align}
Z_{\pm}^{\textup{ph}}= \dfrac{1}{2}\left(1 \pm \dfrac{\Omega^2_+ + \Omega^2_- - 2 \Omega_s^2}{\Omega^2_+ - \Omega^2_-}\right).
\end{align}

The form of the xc energy of Eq.~\eqref{Exc-RPA-Rabi} has a clear physical structure of the zero-point energy, which is typical for RPA \cite{Dobson2005,Furche2008}. The second line in Eq.~\eqref{Exc-RPA-Rabi} can indeed be understood as the interaction-induced change of the zero-point energy of the bosonic electron-photon excitations. 

By inserting $\Omega_{\pm}$ of Eq.~\eqref{eqn:polariton_requenc_one_dimer} into Eq.~\eqref{Exc-RPA-Rabi} we obtain the following compact expression for the photon RPA xc energy, 
\begin{multline}
\!\!E_{xc}^{\rm RPA}\!\!=\!\!\frac{1}{2}\!\left[\!\sqrt{(\omega_{\textup{ph}}+\Omega_s)^2+\lambda^2Z_s^{\textup{ex}}}\!-(\omega_{\textup{ph}}+\Omega_s)\right]\!.\!\!\!  \label{Exc-RPA-Rabi-final}
\end{multline}
As the KS excitation frequency $\Omega_s=2T/\sqrt{1-n^2}$, and the corresponding KS oscillator strength $Z_s^{\rm ex}=4T\sqrt{1-n^2}$ are known as function of $n$, the above equation defines $E_{xc}^{\rm RPA}(n)$ as a simple and explicit function of the density.
\begin{figure}[t!]
\centering
\includegraphics[width=1\linewidth]{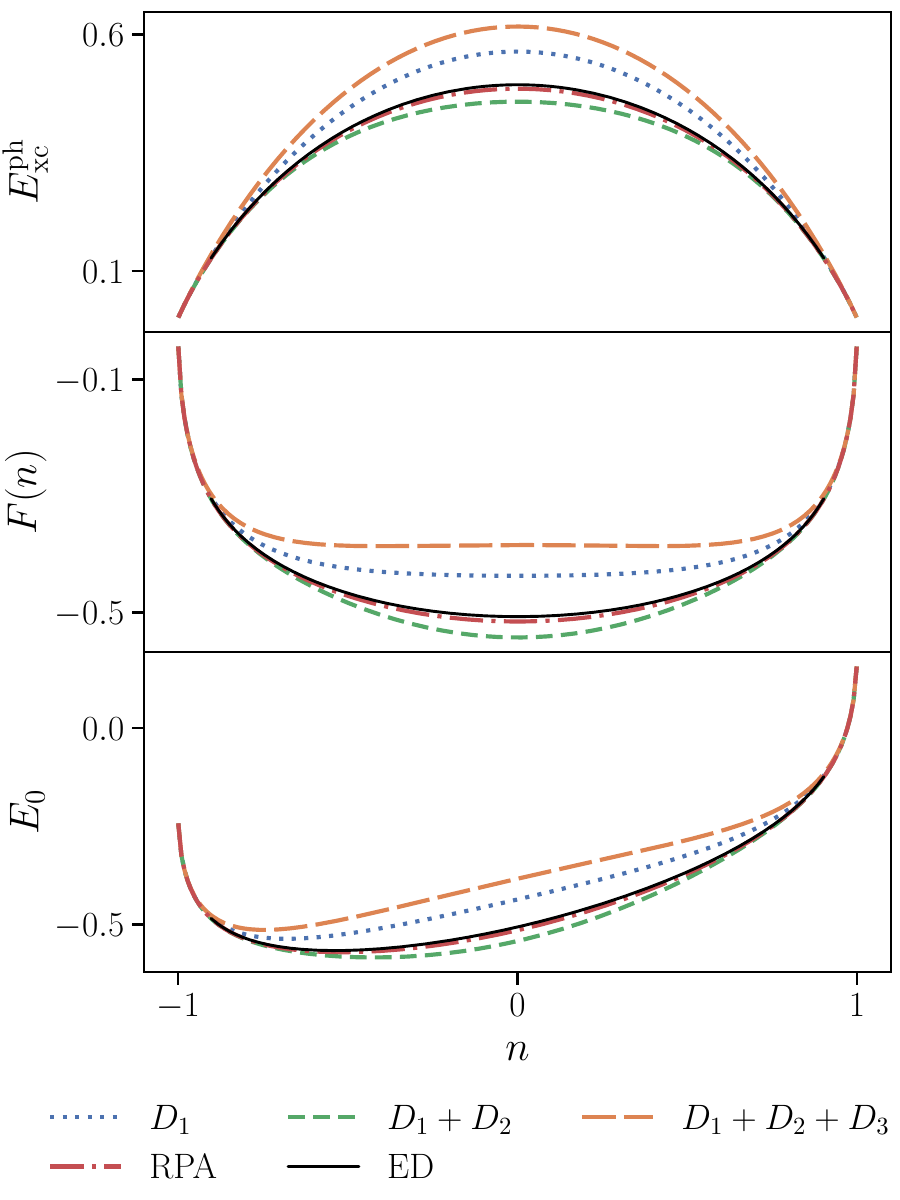}
    \caption{Different contributions to the energy of one dimer in the cavity: xc energy $E_{xc}^\text{ph}$ (upper panel), Hohenberg-Kohn energy $F=T_s+E_{xc}^\text{ph}$ (middle panel), and the full ground state energy $E_0$ (bottom panel). We show approximate functionals based on the finite order diagrams $D_{1,2,3}$ presented in Fig~\ref{fig:basic_diag}, and the full RPA series. The results of exact diagonalization (ED) serve here as a benchmark. Parameters of the system: $\lambda = 1.5$, $\omega_{\textup{ph}} = 1$, $T = 1$. \label{fig:1_dim_energy_as_n}}
\end{figure}
\begin{figure}[t!]
\centering
\includegraphics[width=1\linewidth]{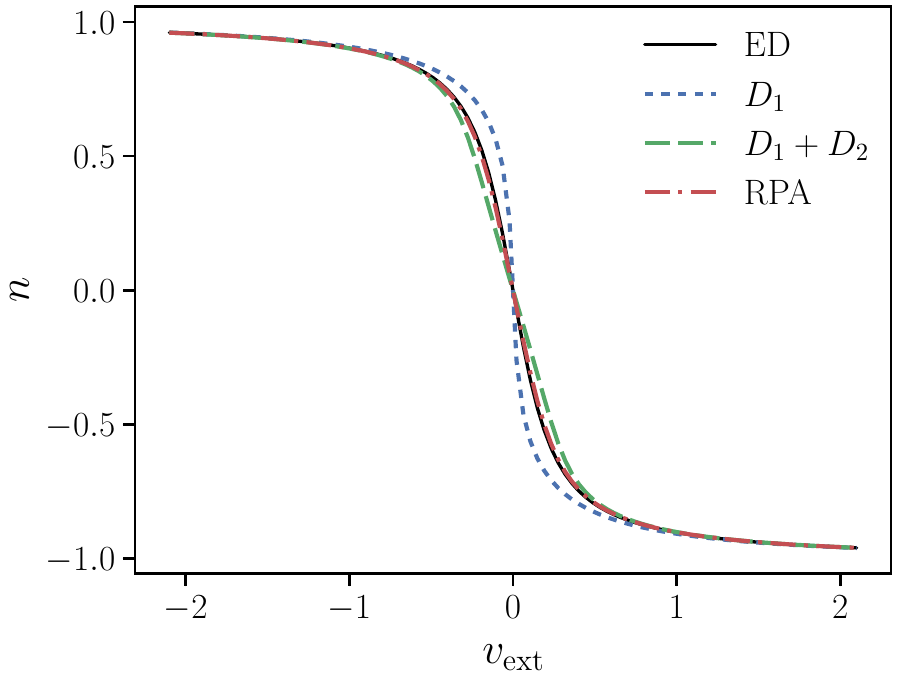}
    \caption{Dependencies of density $n$ on external potential $v_\text{ext}$. ED denotes exact diagonalization. Parameters of the system: $\lambda = 1.3$, $\omega_{\textup{ph}} = 1$, $T = 1$. \label{fig:dn_on_vext_N=1}}
\end{figure}

To assess the quality of the approximate QEDFT functionals for the Rabi ($N=1$ Dicke) model we also compute the xc energy from the exact diagonalization of the Hamiltonian Eq.~(\ref{hamiltonian_rabi}). The exact diagonalization xc energy $E_{xc}^{\rm ED}$ for the many-body ground state $\Psi_0$ is calculated directly from the definition of Eq.~\eqref{Exc-def}, as the difference of $E_0^{\textup{ED}}=\mel{\Psi_0}{\hat H}{\Psi_0}$ and the energy of non-interacting KS system at the same density $n = \mel{\Psi_0}{\hat\sigma_z}{\Psi_0}$,
\begin{align}\label{eqn:XC_energy_ED}
    E_{\textup{xc}}^{\textup{ED}} = E_0^{\textup{ED}} + T\sqrt{1-n^2} - \dfrac{\omega_{\textup{ph}}}{2} - v_{\textup{ext}}n.
\end{align}
To find $E_{\textup{xc}}^{\textup{ED}}$ as function of $n$ one should vary the ground state density $n$. In practice, by calculating the ground state wave function $\Psi_0$ as a function of the external potential $v_{\textup{ext}}$, we define the map $v_{\textup{ext}}\mapsto\{E_{xc}^{\rm ED},n\}$, from which the function $E_{\textup{xc}}^{\textup{ED}}(n)$ is reconstructed. 

In Fig.~\ref{fig:1_dim_energy_as_n} we show different contributions to the energy as functions of the density, namely, the xc energy $E_{xc}^{\rm ph}$ (upper panel), the Hohenberg-Kohn energy $F(n)=T_s + E_{xc}^{\rm ph}$, and the full ground state energy $E_0= F(n) + v_{\textup{ext}}n$. Here, we compare the exact energies with approximate QEDFT energy functionals, at the level of (i) the first order OEP with $E_{xc}^{\rm 1ph}=D_1$, (ii) the full second order OEP corresponding to $E_{xc}^{\rm 2ph}=D_1+D_2+D_3$, (iii) the second order truncated RPA with $E_{xc}^{\rm 2RPA}=D_1+D_2$, and (iv) the full photon RPA defined by Eq.~\eqref{Exc-RPA-Rabi-final}. The corresponding dependencies of the density $n$ on the external potential, obtained by minimizing $E_0= F(n) + v_{\textup{ext}}n$ at fixed $v_{\textup{ext}}$, are presented in Fig.~\ref{fig:dn_on_vext_N=1}. We have chosen a fairly large coupling constant of $\lambda=1.5$, which in the quantum optics tradition would be classified as a deep strong coupling. In our context, the most important point is that in this interaction regime the strict perturbative approaches are not applicable and the difference between approximations becomes obvious.

One can clearly see from Fig.~\ref{fig:1_dim_energy_as_n} that within the first order OEP, the energy does not look correct even qualitatively.  For $\lambda > 1$ this approximation produces a double-well Hohenberg-Kohn functional $F(n)$, while physically there should a single minimum at $n=0$, which we indeed see from the exact solution. This qualitative error is corrected in the second order RPA, that is, by adding the second order RPA-type diagram $D_2$. However, in the full second order OEP, which also includes the second order exchange, the energy functional worsens considerably, becoming even worse than in the first order OEP. The better performance of the "incomplete" second order OEP is not so surprising because we are using the perturbative constructions in a selfconsistent way beyond the strict validity of the perturbation theory. 

Apparently, the best QEDFT energy functional for the Rabi model is obtained within the full photon RPA. The RPA curves in Fig.~\ref{fig:1_dim_energy_as_n} are essentially on top of the results obtained from the exact diagonalization, with the accuracy of a few percent. It is remarkable that the simple function Eq.~\eqref{Exc-RPA-Rabi-final} provides so excellent analytic fit of quite expensive numerical solution of the full many-body electron-photon problem in the deep strong coupling regime. Parametrically the RPA diagrams become dominant in the limit of large $N$, however we see that the photon RPA functional works very well even for $N=1$, at least if the coupling strength is not anomalously large.

Finally, we note that the density distribution shown on Fig.~\ref{fig:dn_on_vext_N=1} is much less sensitive to the quality of the xc density functional. While the differences of $E_{xc}^{\rm ph}(n)$ for different approximations are very well visible, they are much less pronounced in the dependence of the density on the external potential. All functions $n(v_{\textup{ext}})$ in Fig.~\ref{fig:dn_on_vext_N=1} look quite reasonable, in spite the Hohenberg-Kohn functionals in Fig.~\ref{fig:1_dim_energy_as_n} sometimes differ qualitatively. In the next section, we will see that the differences in the performance of different approximations become much more pronounced in the many-dimer Dicke model, that is when several dimers are present in the cavity, being collectively coupled to the photon mode. 

\subsection{General case: Many-dimer Dicke model}

\subsubsection{$N$ equivalent dimers coupled to the cavity mode}
Now we consider a group of $N$ equivalent dimers in the cavity, biased by the same external potential $v_{\textup{ext}}$. In this case, the Hamiltonian of Dicke model, Eq.~\eqref{H-Dicke-def}, reduces to the form,
\begin{multline}
    \hat{H}=\sum\limits_{i=1}^{N}\left[-T\hat{\sigma}_x^{i}+\left[\sqrt{\frac{\omega_{\textup{ph}}}{2}}\lambda(\hat{a}+\hat{a}^{\dagger})+v_{\textup{ext}}\right]\hat{\sigma_{z}}^{i}+\frac{\lambda^2}{2}\right]\quad \\
    +\omega_{\textup{ph}}(\hat{a}^{\dagger}\hat{a}^{}+1/2) +\sum\limits_{i=1}^{N-1}\sum\limits_{j=i+1}^{N}\lambda^2\hat{\sigma_{z}}^{i}\hat{\sigma}_{z}^{j}.\label{hamiltonian_rabi_many_dimers_equiv}
\end{multline}
where we used the second quantized representation of Eq.~\eqref{photon-operators} for the photon canonocal variables. Apparently, the local densities $n=\langle\sigma_z^i\rangle$ are identical for all dimers. The same is true for one-particle energies attributed to separate dimers in the KS system, so that the kinetic energy $T_s$ of the KS system and the energy of external potential scale with $N$ thus depending extensively on the number of dimers. The total ground state energy can then be written in terms of the density $n$ of single dimer as follows,
\begin{align}\label{eqn:full_energy_RPA_many_dimers}
&E_0=-NT\sqrt{1-n^2}+Nv_{\textup{ext}}n + E_{\textup{xc}}^{\textup{ph}}(n)+\dfrac{\omega_{\textup{ph}}}{2}.
\end{align} 
Since the dimers are coupled to the same photon mode their behavior is expected to be correlated. In particular, the ground state density in a given dimer should depend on the presence of the other dimers in the cavity. In the rest of this section we study the importance of such collective effects in the ground state, and to which extent they can be captured by different approximations for the xc energy functional.

Formally the above mentioned correlations are reflected in the dependence of $E_{xc}^{\rm ph}$ on the number $N$ of dimers in the cavity. Within our diagramatic approach, a given diagramm for xc energy scales as $N^l$, where $l$ is the number of electronic loops. For example, we have $E_{xc}^{\rm 1ph}=ND_1$ for the first order OEP,  $E_{xc}^{\rm 2RPA}=ND_1+N^2D_2$ for the second order truncated RPA, $E_{xc}^{\rm 2ph}=ND_1+N^2D_2+ND_3$ for the full second order OEP, where $D_1$, $D_2$, and $D_3$ are given by Eqs.~\eqref{D1}-\eqref{D3} in Appendix~A. Therefore, the first order OEP scales extensively with $N$, which implies that the dimers do not feel each other, and the physically expected correlations are totally missing in this approximation. This fictitious extensivity is violated already at the second order by the RPA diagram (the second diagram on Fig.~\ref{fig:basic_diag}) which scales as $N^2$. 

Apparently, the terms of higher order in the electron-photon coupling will generate contributions with higher powers in $N$. An important observation is that, within a given order of the diagrammatic perturbation theory for $E_{xc}^{\rm ph}$, the RPA diagram has the maximal number of fermionic loops thus corresponding to the highest power of $N$. This suggests that the quality of the photon RPA functional should further improve with the increase of the number of dimers. The explicit form of the RPA xc energy for the Dicke model is immediately obtained from Eqs.~\eqref{Exc-RPA-Rabi} and \eqref{Exc-RPA-Rabi-final} by noticing that the KS polarizability is additive, and for $N$ dimers it simply reads $\Pi_s^N(\omega)=N\Pi_s(\omega)$. Therefore, the final RPA xc energy is obtained by replacing $\lambda^2\mapsto N\lambda^2$ in the corresponding result for $N=1$,
\begin{multline}
E^{\textup{RPA}}_{xc}=\int\limits_{0}^{\infty}\dfrac{d\omega}{2\pi}\textup{ln}\left[1-N\lambda^2D_{\textup{ph}}(i\omega)\Pi_s(i\omega)\right]\\=\frac{1}{2}\left[\sqrt{\lambda^2 N Z_s^{\textup{ex}}+(\omega_{\textup{ph}}+\Omega_s)^2}-(\omega_{\textup{ph}}+\Omega_s)\right].
\label{Exc-RPA-Dicke}
\end{multline}
Physically this energy can be interpreted as the interaction correction to the zero-point energy of bright (dipole active) collective polaritons with the following frequencies,
\begin{align} \label{OmegaNpm}
\Omega_{N\pm}&=\dfrac{1}{2}\left[\sqrt{(\omega_{\textup{ph}}+\Omega_s)^2+\lambda^2 N Z_s^{\textup{ex}}}\right. \\ & \quad\qquad\qquad\qquad\pm \left.\sqrt{(\omega_{\textup{ph}}-\Omega_s)^2+\lambda^2 N Z_s^{\textup{ex}}}\right].\nonumber 
\end{align}
The difference between these frequencies correspond to the collective Rabi splitting that has a familiar structure,
\begin{equation}
    \label{splittingN}
    \Omega_R = \sqrt{(\omega_{\textup{ph}}-\Omega_s)^2+\lambda^2 N Z_s^{\textup{ex}}},
\end{equation}
while their sum determines the xc energy of Eq.~\eqref{Exc-RPA-Dicke}. For sufficiently large $N$ the frequency $\Omega_{N+}$ of the upper polaritonic excitation, and the Rabi splitting $\Omega_R$ grow as $\sqrt{N}$, which is commonly interpreted as a regime of ultrastrong collective coupling [?Refs?]. On the other hand, this automatically assumes that the xc energy also scales with $\sqrt{N}$ when the number of molecules in the cavity increases. Such sub-extensive scaling implies decreasing of the xc energy per dimer as $1/\sqrt{N}$, which means that the relative contribution of the xc effects to the properties of the ground state becomes progressively smaller when we go to the ultrastrong collective coupling. We will return to this seemingly counter-intuitive point in a moment, after discussing the performance of different approximate xc functionals for the many-dimer Dicke model.
\begin{figure}[t!]
\centering
\includegraphics[width=1\linewidth]{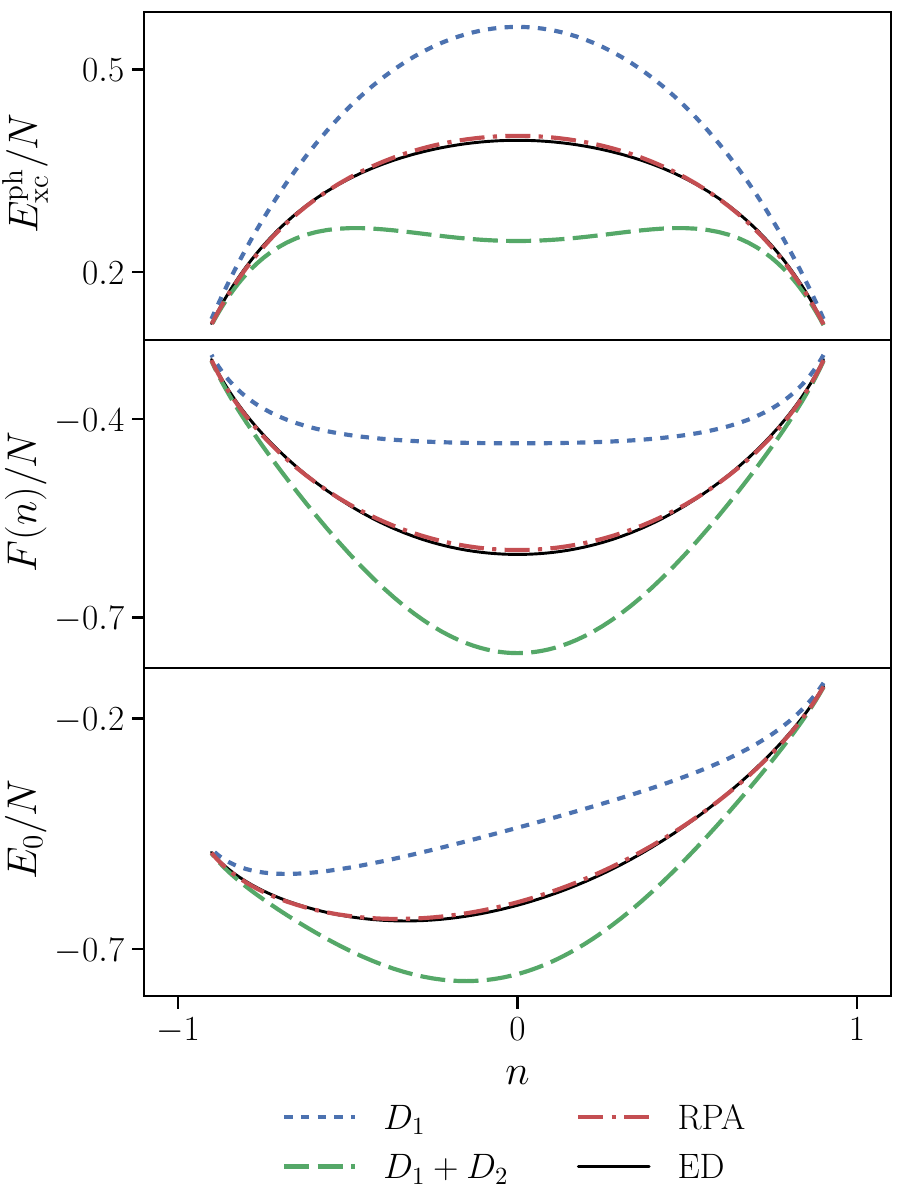}
    \caption{Different contributions to the energy per dimer for a system of $N$ equivalent dimers in the cavity as functions of the one-dimer density $n$: xc energy $E_{xc}^\text{ph}$ (upper panel), Hohenberg-Kohn energy $F=T_s+E_{xc}^\text{ph}$ (middle panel), and the full ground state energy $E_0$ (bottom panel), calculated in different approximation. The results of exact diagonalization (ED) serve here as a benchmark. Parameters of the system: $N=3$, $\lambda = 1.5$, $\omega_{\textup{ph}} = 1$, $T = 1$.}\label{fig:Ex_diff_many_dim}
\end{figure}
\begin{figure}[t!]
\centering
\includegraphics[width=1\linewidth]{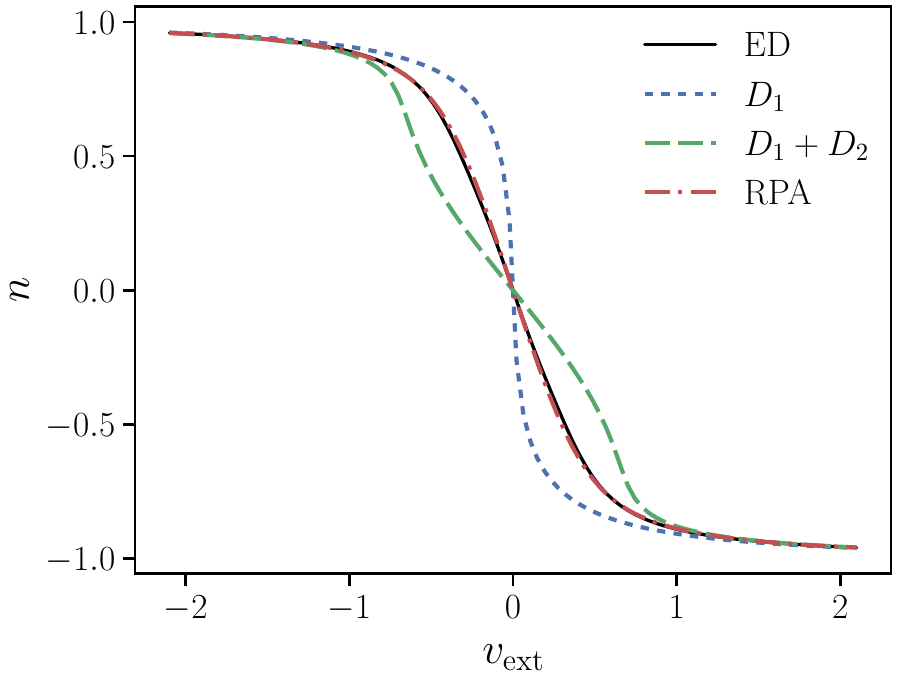}
    \caption{Dependency of the one-dimer density $n$ on external potential $v_\text{ext}$ for the $N$-dimer Dicke model. Parameters of the system: $N = 3$, $\lambda = 1.3$, $\omega_{\textup{ph}} = 1$, $T = 1$. \label{fig:dn_on_vext}}
\end{figure} 

In Fig.~\ref{fig:Ex_diff_many_dim} we show the xc energy, Hohenberg-Kohn energy, and the total ground state energy per dimer for $N=3$ and the same coupling strength as in Fig.~\ref{fig:1_dim_energy_as_n}. Here we compare the results of three approximations, the first order OEP ($D_1$), the second order RPA ($D_1+D_2$), and the full photon RPA, with those obtained from the exact diagonalization (ED). It is clear that already for $N=3$ the OEP schemes based on a finite order perturbation theory produce unreliable energetics (we do not show the results of the full second order OEP which look even worse). In contrast, the full nonperturbative RPA, as expected, approximate the exact energy functionals almost perfectly. The ground state density of a given dimer in Fig.~\ref{fig:dn_on_vext} confirms this trend. Despite the density is less sensitive to the quality of approximation, the deficiency of the finite order OEP functionals becomes obvious in the case of few a dimers in the cavity. We note that the first order produces at least a qualitatively reasonable density without unphysical inflections. However, it strongly overestimates the density compared the results of the exact diagonalization. In contrast, the RPA density and the exact one are essentially indistinguishable.    

\begin{figure}[t!]
\centering
\includegraphics[width=1\linewidth]{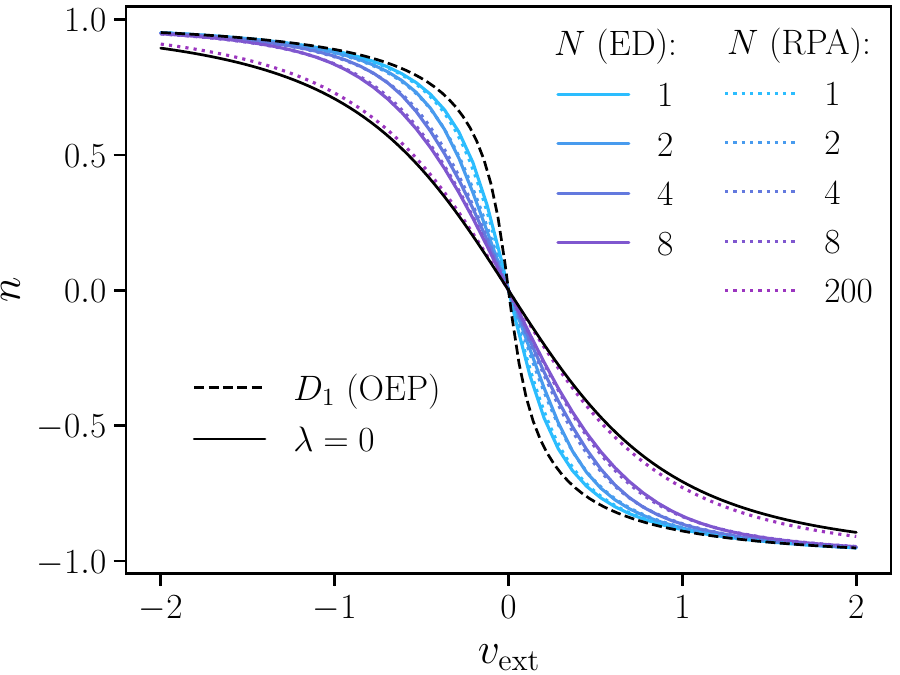}
    \caption{Dependencies of the one-dimer density $n$ on the external potential for different numbers of dimers $N$ calculated within RPA and the exact diagonalization. Parameters of the system: $\omega_{\textup{ph}} = 1$, $T = 1$, $\lambda=1.2$. \label{fig:n_for_many_dimers_comp}}
\end{figure}

The systematic overestimation of the effect of the cavity by the first order OEP is related to its wrong scaling with $N$ and missing collective effects. To illustrate this point, in Fig.~\ref{fig:n_for_many_dimers_comp} we present the dependence of the density in a given dimer on the external potential for different number of dimers in the cavity. As we can see, the function $n(v_{\textup{ext}})$ corresponding to the first order OEP can be considered as an upper bound for the exact/RPA densities. It reasonably agrees with the exact $n(v_{\textup{ext}})$ at $N=1$, but becomes progressively worse for lager $N$. While the first order OED does not depend on $N$, the exact/RPA treatment shows that the xc correction to the density gets smaller with the increase of the number of dimers. In fact, the lower bound for the density in Fig.~\ref{fig:n_for_many_dimers_comp} is provided by the curve for the decoupled system with $\lambda=0$. For the considered coupling strength $\lambda\gtrsim 1$, this lower bound is reached at $N>100$, which we have checked explicitly within RPA as the exact diagonalization becomes too costly for $N>10$.

Apparently in the case of large $N$, corresponding to the ultrastrong collective coupling, the cavity photon mode produces a progressively smaller effect on the electron ground state density. This can be understood as follows. In the DFT language, the photon xc correction to the density in a dimer is controlled by the xc potential $v_{xc}(n)$, which is given by density derivative of the xc energy per dimer,
\begin{equation}
    \label{vxc}
v_{xc}=\frac{1}{N}\frac{\partial E_{xc}^{\rm ph}}{\partial n}.
\end{equation}
As in the limit of large $N$ the xc energy is expected to scale sub-extensively $\sim\sqrt{N}$, the xc potential decreases as $v_{xc}\sim 1/\sqrt{N}$ thus leading to the smaller xc shift of the density. 

This behavior can be explained more physically in terms of the many-body perturbation theory. Let us consider the self energy of the electron induced by the virtual excitation of the dressed photon propagating through a polarizable system of dimers. In the case of $N$ dimers, the propagator of the physical dressed photon, by analogy with Eq.~\eqref{eqn:dressed_prop}, reads as follows,
\begin{equation}
    \label{photon-propag-N}
 \mathcal{D}_N(\omega)  =-\omega^2\left(\dfrac{Z^{\textup{ph}}_{N+}}{\omega^2-\Omega_{N+}^2}+\dfrac{Z^{\textup{ph}}_{N-}}{\omega^2-\Omega_{N-}^2}\right),
\end{equation}
where the frequencies of the bright collective polaritons are given by Eq.~\eqref{OmegaNpm}, and the strengths of the corresponding poles in the propagator is defined as,
\begin{align}
\!\!Z_{N\pm}^{\textup{ph}}\!=\! \dfrac{1}{2}\!\left[1 \! \pm \dfrac{\lambda^2 N Z_s^{\textup{ex}} +\omega_{\textup{ph}}^2-\Omega_{s}^2}{\sqrt{(\lambda^2 N Z_s^{\textup{ex}}\!\!+\Omega_s^2\!+\!\omega_{\textup{ph}})^2\!-4\omega^2_{\textup{ph}}\Omega_s^{2}}}\right].
\end{align}
The virtual excitation of the polartonic excitation contributing to the physical photon is the physical origin of cavity induced electron self energy.
In the limit of a large $N$, the photon propagator is dominated by the upper polarition as its weight $Z_{N+}^{\textup{ph}}$ tends to unity. However, because the frequency $\Omega_{N+}$ grows as $\sqrt{N}$ the excitation of this mode becomes expensive, leading to the $O(1/\sqrt{N})$ correction to the self energy. On the other hand, the frequency of the lower polariton $\Omega_{N-}$ decreases as $1/N$, and it could be efficiently excited, but this is compensated by its small weight factor $Z_{N-}^{\textup{ph}}$. This argument shows that in the regime ultra strong collective coupling the huge modification of the polaritonic excitations and very large Rabi splitting imply the weak influence of the cavity on the ground state properties of the electronic subsystem. It is worth noting that the effective decoupling of the electronic subsystem in the large $N$ limit is closely to the so-called no-go theorems preventing superradiace in the thermodynamic limit \cite{Andolina2019}.

Our results clearly demonstrate a very good performance of the photon RPA functional for QEDFT of the Dicke model. The RPA xc energy in Eq.~\eqref{Exc-RPA-Dicke} provides a simple, but excellent analytic fit of the exact xc energy obtained from a heavy numerical diagonalization of the full many-body Hamiltonian in a strongly nonperturbative regime. Now we analyze quantitatively the accuracy of this approximation for different number of dimers in the cavity.

\begin{figure}[t!]
\centering
\includegraphics[width=1\linewidth]{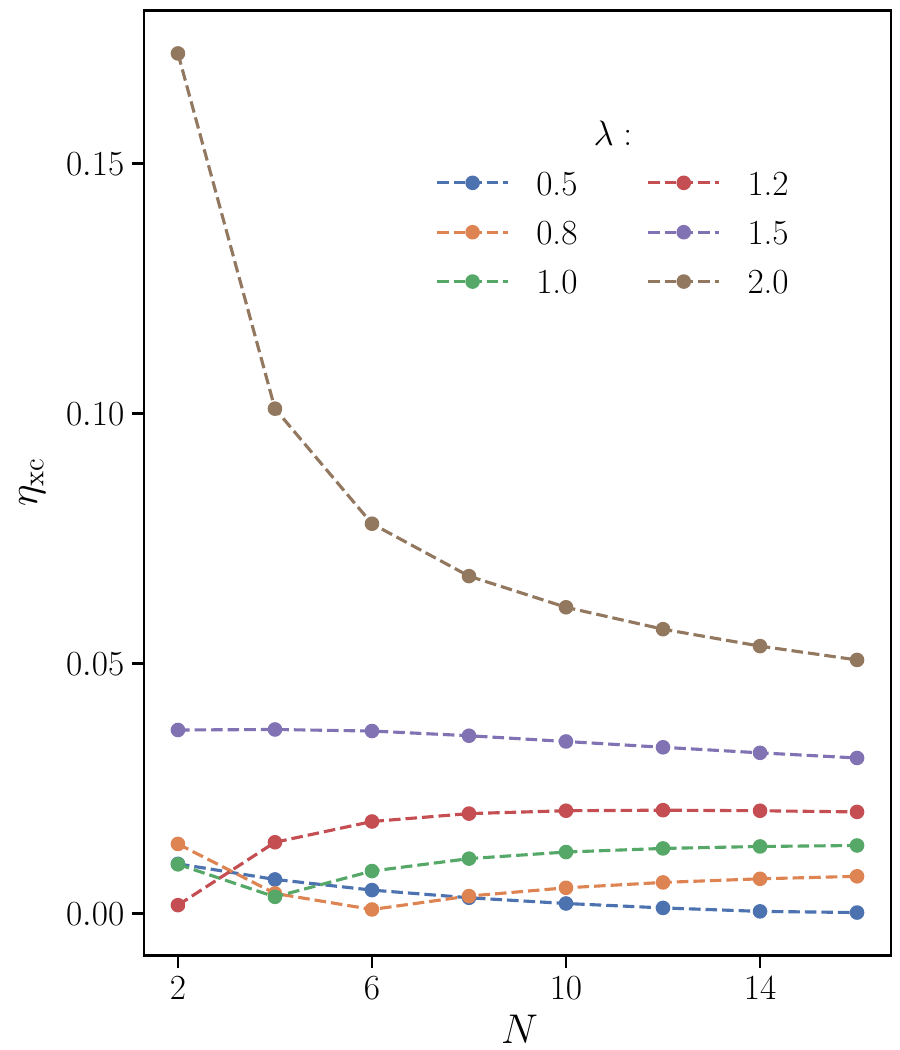}
    \caption{The relative error $\eta_\textup{xc}$ of the RPA xc energy with respect to the exact results, depending on the number of dimers $N$. Different plots correspond to different coupling constant $\lambda$. Parameters of the system: $\omega_{\textup{ph}} = 1$, $T = 1$. \label{fig:Ex_diff_Ndim}}
\end{figure}

\begin{figure}[t]
\centering
\includegraphics[width=1\linewidth]{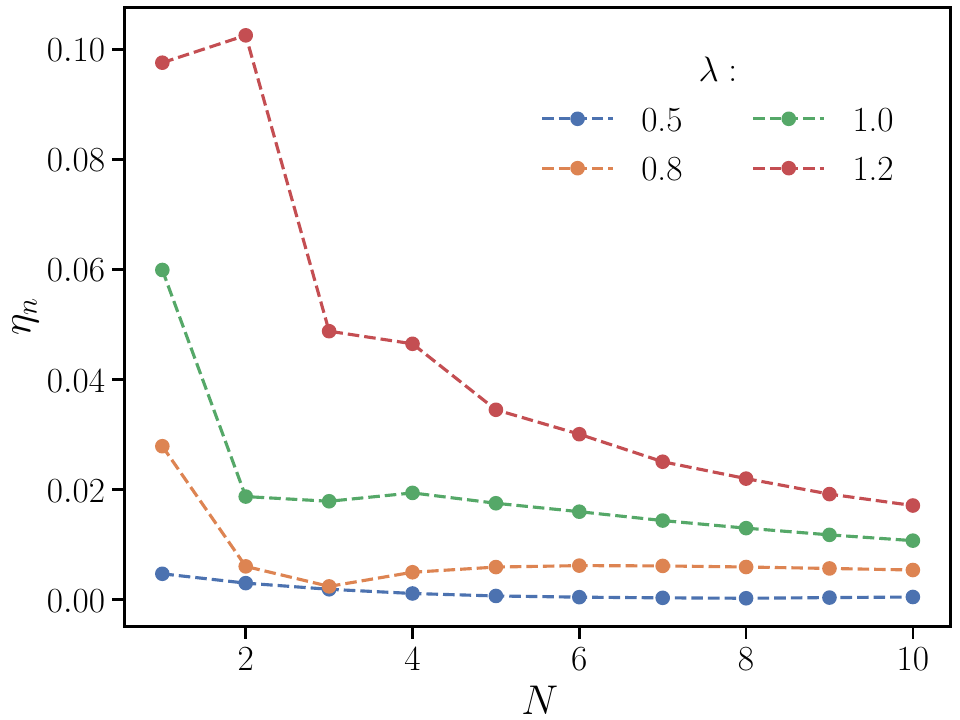}
    \caption{The density error $\eta_n$ as a function of number of dimers $N$ for different $\lambda$. Parameters of the system: $\omega_{\textup{ph}} = 1$, $T = 1$. \label{fig:dn_diff_on_N_relative}}
\end{figure}

It is convenient to quantify the accuracy of the RPA xc energy by the maximal relative error $\eta_{\rm xc}$ of $E_{xc}^\textup{RPA}$ with respect to exact diagonalization xc energy $E_{xc}^\textup{ED}$, 
\begin{align}
\eta_{\textup{xc}}\!=\!\norm{\frac{E_\textup{xc}^\textup{ED}\!-\!E_\textup{xc}^\textup{RPA}\!}{E_\textup{xc}^\textup{ED}}}_\infty\!\!\!\!\!=\underset{n}{\max} \qty|\frac{E_\textup{xc}^\textup{ED}(n) - E_\textup{xc}^\textup{RPA}(n)}{E_\textup{xc}^\textup{ED}(n)}|.
\end{align}
In fact, for all $N$ the maximal error is achieved at $n=0$ that corresponds to symmetric dimers. Similarly, we characterize the accuracy of computing the density from the RPA, by the relative density error,
\begin{align}
    \eta_n &=\norm{\frac{n^\textup{ED}-n^\textup{RPA}}{n^\textup{ED}}}_\infty= \nonumber\\ 
     &\qquad\qquad\quad= \underset{v_\textup{ext}}{\max}\qty|\frac{n^\textup{ED}(v_\textup{ext}) - n^\textup{RPA}(v_\textup{ext})}{n^\textup{ED}(v_\textup{ext})}|,
\end{align}
where $n^\textup{RPA}(v_\text{ext})$ and $n^\textup{ED}(v_\text{ext})$ are the densities calculated from RPA functional within QEDFT and from the exact diagonalization, respectively.

The dependence of $\eta_\text{xc}$ on the number of dimers $N$ is given in Fig.~\ref{fig:Ex_diff_Ndim}. It shows that for the coupling strength $\lambda\sim 1$ the error of the RPA xc energy gets larger for larger couplings, but remains small, typically at the level of few percent. For larger values of $\lambda$ we observe the expected increase of accuracy (decrease of the relative error) with increasing the number of dimers. The density error presented in  Fig.~\ref{fig:dn_diff_on_N_relative} shows the same trends with the same level of error. 

An interesting observation is that in the limit of large $N$ the errors in Figs.~\ref{fig:Ex_diff_Ndim}-\ref{fig:dn_diff_on_N_relative}, while becoming smaller, do not go to zero, but saturate to some small finite value. In other words, we see a numerical evidence that RPA, being quite accurate, is not asymptotically exact at $N\gg 1$. The reason is the adopted way of taking the limit of large $N$ at fixed coupling constant $\lambda$, which indicates an interesting subtlety in defining the thermodynamic limit in the cavity QED context. 

Physically, $\lambda\sim 1/\sqrt{V_\text{m}}$ is inversely proportional to the square root of the mode/cavity volume. Hence, by increasing $N$ at fixed $\lambda$ we put more molecules inside a given cavity. In this absolutely physical setup, for a given order of the perturbation theory, the RPA diagram carries the largest possible power of $N$ thus dominating over all other contributions of the same order. However, the same power of $N$ can also come from the higher order non-RPA diagrams. In a non-perturbative regime with $\lambda > 1$ such contributions formally can not be ignored. Our exact diagonalization results show that, as a matter of fact, the post-RPA corrections are small, but it is clear that in this setup RPA is never exact. The observed saturation of the xc energy error indicates that the scaling of the exact xc energy $E_\text{xc}^\text{ph}\sim\sqrt{N}$ is probably the same as in RPA, but the coefficient is slightly different.

\begin{figure}[t!]
\centering
\includegraphics[width=1\linewidth]{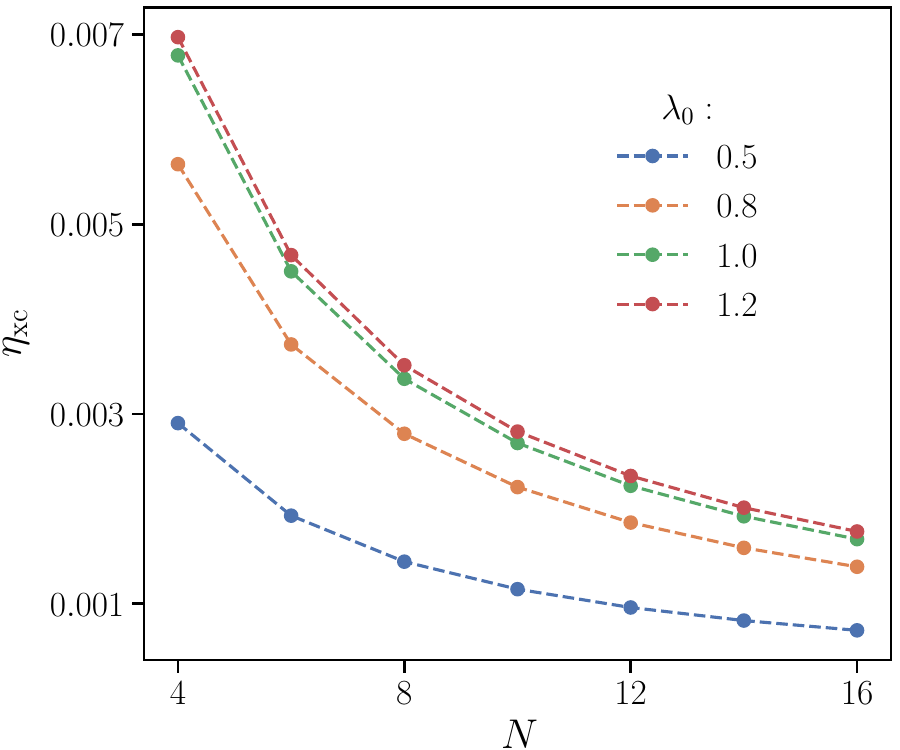}
    \caption{The relative error of the RPA xc energy as a function of $N$. Here, the change of $N$ is accompanied  with rescaling of the coupling $\lambda=\boldsymbol{\lambda_0/\sqrt{N}}$. In all cases, the parameters of the system are: $\omega_{\textup{ph}} = 1$, $T = 1$. \label{fig:Ex_diff_Ndim_norm}}
\end{figure}

Alternatively, we can increase the number of molecules $N$ in the cavity while keeping fixed their density $N/V_\text{m}$ inside the mode volume. This assumes increasing the mode volume $V_\text{m}\sim N$, and therefore rescaling the coupling constant as $\lambda =\lambda_0/\sqrt{N}$. We note that rescaling of the coupling constant with $N$ is a common way of defining the large $N$ expansion in the quantum field theory, see for example Ref.~\cite{Altland-book}. In the cavity QED context, it can also be understood as fixing the so-called collective coupling strength $\lambda\sqrt{N} = \lambda_0$. If the limit of large $N$ is taken at fixed $\lambda_0$, each photon propagator in a diagram brings the factor of $1/N$, which makes RPA asymptotically exact in the limit of large $N$. In fact, the RPA energy diagrams of all orders become $N$-independent, while all other contributions appear with negative powers of $N$. The xc energy error $\eta_\text{xc}$ in this setup is presented in Fig.~\ref{fig:Ex_diff_Ndim_norm}. Our numerical calculations of the exact xc energy indeed show that the error of RPA energy converges to zero at large $N$. In our case, it reaches fractions of percent at $N\gtrsim 10$.

\subsubsection{Arbitrary set of dimer groups}

In the previous section, we have demonstrated that QEDFT based on RPA works extremely well for any $N$ and independently of a specific way of controlling the number of dimers in the cavity. In this demonstration, we used the simplest version of the Dicke model, in which  equivalent dimers are all biased by the same external potential. It is natural to expect that this technical simplification should not be critical for the quality of RPA.

In this section, we discuss QEDFT for the most general gauge invariant Dicke model, and derive the explicit form of the corresponding RPA energy functional. Specifically, we consider an arbitrary number $M_{\textup{d}}$ of different groups of equivalent dimers, coupled to a photon mode. The $j^{\textup{th}}$ group consists of $N_j$ dimers, so that we have $N=\sum_{j=1}^{M_\text{d}}N_j$ dimers in total. The dimers within $j^{\textup{th}}$ group are subject to the same external field $v_{\textup{ext}^j}$, have coupling constant $\lambda_j$, and the intra-dimer hopping $T_j$. The corresponding generalization of the Dicke Hamiltonian Eq.~\eqref{H-Dicke-def} takes the form,
\begin{multline}
   \!\!\!\! \hat{H}=\sum\limits_{j=1}^{M_{\textup{d}}}\sum\limits_{i=1}^{N_j}\left[-T_j\hat{\sigma}_x^{ij}+\left[\sqrt{\frac{\omega_{\textup{ph}}}{2}}\lambda_j(\hat{a}+\hat{a}^{\dagger})+v_{\textup{ext},j}\right]\hat{\sigma_{z}}^{ij}\right]\\
\!\!+\sum\limits_{j_1=1}^{M_{\textup{d}}}\!\sum\limits_{j_2=1}^{M_{\textup{d}}}\sum\limits_{i_1=1}^{N_{j_1}}\sum\limits_{i_2=1}^{N_{j_2}}\lambda_{j_1}\lambda_{j_2}\hat{\sigma_{z}}^{i_1j_1}\hat{\sigma}_{z}^{i_2j_2}+\omega_{\textup{ph}}(\hat{a}^{\dagger}\hat{a}^{}+1/2).\label{hamiltonian_rabi_many_dimers_dif}
\end{multline}

By construction, the dimers in the same group have identical densities. Therefore, QEDFT for this system operates with $M_\text{d}$ basic density variables $n_j,\quad j=1\dots M_\text{d}$. Within the KS formulation of the theory, the ground state energy as a function of the density reads as follows,
\begin{multline}
\!\! E_0=\sum\limits_{i=1}^{M_{\textup{d}}}N_i\big[T_s(n_i) + v_{\textup{ext}^i}n_i\big] +\dfrac{\omega_{\textup{ph}}}{2} + E_{\textup{xc}}^\text{ph}\left(\{n_i\}\right),
\end{multline}
where $T_s(n_i)=-T_i \sqrt{1-n_i^2}$ is the KS kinetic energy of a dimer in $i$th group. 

The RPA xc energy in this system is given by the obvious generalization of the expression for the single group Dicke model,
\begin{multline} \label{Exc-RPA-general-Dicke-def}
E_\textup{xc}^{\rm RPA}
=\int\limits_{0}^{\infty}\dfrac{d\omega}{2\pi}\textup{ln}\left[1-\sum\limits_{j=1}^{M_{\textup{d}}}\lambda_j^2  N_j D_{\textup{ph}}(i\omega)\Pi_{s j}(i\omega)\right],
\end{multline}
where the KS polarizability $\Pi_{sj}(\omega)$ of a dimer from $j^{\textup{th}}$ group is defined as follows,
\begin{align}
\Pi_{sj}(\omega)=-\dfrac{Z^{\textup{ex}}_{sj}}{\omega^2-\Omega_{sj}^2}.
\end{align}
Here $Z_{sj}^{\textup{ex}}=4 T_j \sqrt{1-n_j^2}$ and $\Omega_{sj}= 2 T_j/\sqrt{1-n_j^2}$ are, respectively, the oscillator strength and excitation frequency of the KS exciton corresponding to a dimer in $j^{\textup{th}}$ group.  
 
Similarly to the single dimer and the singe group cases, after the integration in Eq.~\eqref{Exc-RPA-general-Dicke-def} the xc energy $E^{\rm RPA}_{xc}$ is expressed in terms of the zero-point energy of the bright polaritons. However, in the general case of $M_\text{d}$ inequivalent groups of dimers, there are $M_\text{d}+1$ optically active polaritonic excitations with frequencies $\Omega_\nu$. Apparently, the number of bright polaritons in the interacting system equals to the number of distinct KS excitons (number of dimer groups $M_\text{d}$) plus the number of photon modes (one in our case). The final result for the photon RPA xc energy then takes the following form,
\begin{align} \label{Exc-RPA-Dicke-general-fin}
E^{\rm RPA}_\textup{xc}=\!\dfrac{1}{2}\!\left[\!\sum\limits_{\nu=1}^{M_{\textup{d}}+1}\!\! \Omega_\nu \! - \! \! \Big(\sum\limits_{j=1}^{M_{\textup{d}}}\!\Omega_{sj}  +  \omega_{\textup{ph}} \! \Big)\! \right]\!.
\end{align}
The frequencies $\Omega_\nu$ of bright polaritons correspond to zeroes of the argument of the logarithm in Eq.~\eqref{Exc-RPA-general-Dicke-def}. Alternatively, they can be defined as positive roots of the following function,
\begin{align} \label{PN}
P_{M_{\textup{d}}}(\omega)=\omega_{\textup{ph}}^2 - \omega^2\left(1 - \sum\limits_{j=1}^{M_{\textup{d}}}\lambda_j^2  N_j \Pi_{sj}(\omega)\right).
\end{align}
Obviously, the roots of $P_{M_{\textup{d}}}(\omega)$ also determine the position of poles in the dressed photon propagator. 

Equations \eqref{Exc-RPA-Dicke-general-fin} and \eqref{PN} fully determine the xc energy $E_{xc}^\text{RPA}(n_1,\dots,n_{M_{\rm d}})$ as a function of $M_{\rm d}$ densities. A closed analytic expression for this function can still be found in the case of two groups, $M_{\rm d}=2$. The corresponding formulas are presented in Appendix~B. In general, for $M_{\rm d}>2$ a numerical solution of the polynomial equation $P_{M_{\textup{d}}}(\omega)=0$ is required, which is nonetheless incomparably easier than the full numerical solution of the interacting many-body electron-photon problem defined by the Hamiltonian of Eq.~\eqref{hamiltonian_rabi_many_dimers_dif}.  

\begin{figure}
\centering
\includegraphics[width=1\linewidth]{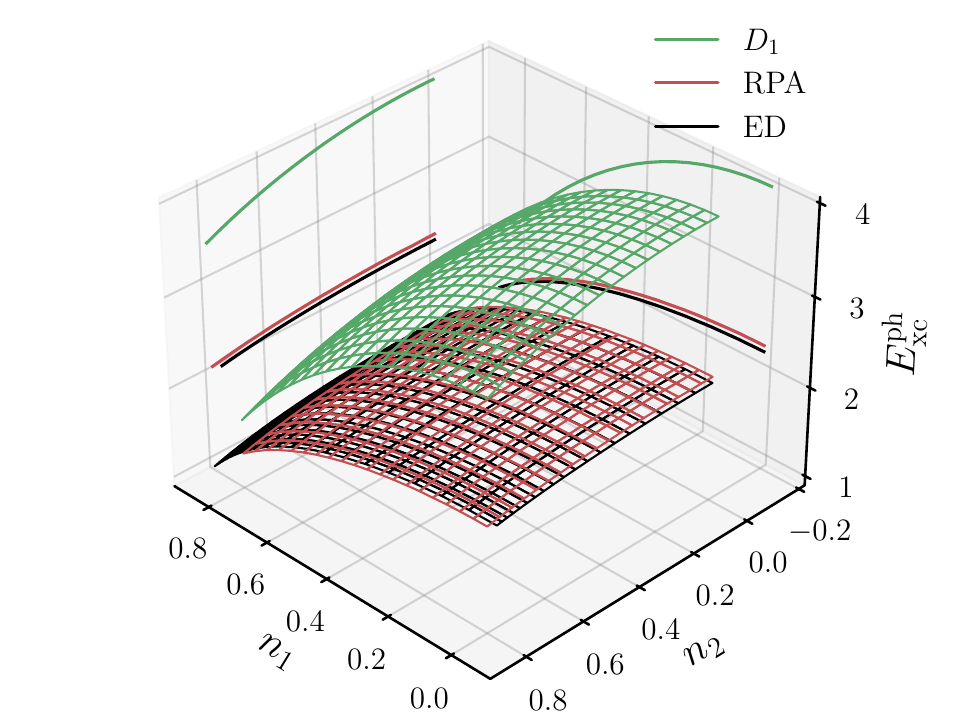}
    \caption{The xc energy as a function of the densities, $n_1$ and $n_2$, for a Dicke model with two groups of dimers: The first order OEP $E_{xc}^{\rm 1ph}$ (green surface), the RPA ex energy $E_{xc}^{\rm RPA}$ (red surface), and the exact diagonalization $E_{xc}^{\rm ED}$ (black surface). Projections along $n_1$ and $n_2$ axes show slices of these surfaces at $n_1 = 0$ and $n_2 = 0$, respectively. The number of dimers, coupling constants and hopping coefficients within each group are: $N_1 = 4$, $N_2 = 3$, $\lambda_1 = 1.5$, $\lambda_2 = 1$, $T_1=T_2=1$. Photon frequency: $\omega_{\textup{ph}} = 1$. \label{fig:Exc_surface}}
\end{figure}

The performance of the photon RPA for the general multi-group Dicke model is illustrated in Fig.~\ref{fig:Exc_surface}. Specifically, we consider a system with $M_\text{d}=2$ which contains four dimers in the first group $N_1=4$, and tree dimers in the second group $N_2=3$. The 3d plots of xc energies $E_{xc}^{\rm ph}(n_1,n_2)$ presented in Fig.~\ref{fig:Exc_surface} show that for this system in a nonperturbative regime with couplings $\lambda_i\gtrsim 1$, the first order OEP denoted as $D_1$ fails dramatically. As we have already discussed, this failure is mostly related to missing collective effects, which leads to the strong overestimation of the cavity induced modifications of the electron density. In contrast, the full photon RPA agrees excellently with the results of exact diagonalization, demonstrating the expected few percent accuracy.

Despite clearly demonstrated excellent performance of the photonic RPA functional, it is worth adding a word of caution. If the coupling constant becomes substantially larger than unity, and the number of dimers is not too large, the errors of RPA grow, see for example the curve corresponding to $\lambda=2$ in the range of small $N$ in Fig.~\ref{fig:Ex_diff_Ndim}. This trend continues with the further increase of the coupling strength. Moreover, for mesoscopic (consistent of few dimers) systems at sufficiently strong coupling, new physical effects, such as a quantized step-like cross-polarizability have been predicted recently  \cite{Kudlis2023arxiv}. Whether these effects can be captured by some regular post-RPA corrections, for example based on the $1/N$ expansion, is an interesting question for the future.

\section{Conclusion}
In conclusion, we presented a comprehensive formulation of ground state QEDFT in application to the generalized Dicke model. In particular, we derived the general exact representation of the xc energy in the KS DFT based on the ACFDT formalism in the case when electrons are coupled to the long wavelength cavity photons. The main emphasis of this work is to adopt the ACFDT representation of the xc energy, and the many-body perturbation theory for constructing explicit approximate xc functionals and carefully test their performance using the Dicke model as a specific simple example.   

The gauge invariant version of the Dicke model can be viewed as a cartoon of a polaritonic chemistry setup in which a collection of diatomic molecules described tight-binding dimers and minimally coupled to the cavity photons. This model is expected to correctly capture the collective nature of the interaction between molecules and the photons, and the importance of this collectivity for the ground state QEDFT is one of the main general conclusions of our work. 

Specifically, for the Dicke model, we derived several OEP-type approximate xc functionals and check their performance against the exact numerical diagonalization of the Dicke Hamiltonian. We analyzed the OEP constructions based on a finite order perturbation theory, similar to the first order one-photon OEP of Ref.~\cite{Pellegrini2015PRL}, and the approximation for xc energy corresponding to the full RPA diagrammatic series. Our results show that in the non-perturbative regime of strong coupling, $\lambda\gtrsim 1$ and any number $N$ of molecules in the cavity, RPA functional works extremely well with an error of about few percent. In contrast, the finite order OEP constructions may fail dramatically, especially for $N>1$, that is, when several molecules are coupled to the same photon mode. The origin of this failure is a wrong scaling with the number of molecules in the cavity. In particular, the first order OEP, being simply proportional to $N$, completely missed collective effects and as a result strongly and systematically overestimates the interaction corrections to the ground state electron density. The physical photon contribution to the xc energy scales at large $N$ sub-extensively $\sim\sqrt{N}$, which is perfectly captured by RPA. 

Apparently, the importance of collective effects influencing the dependence on the number of emitters in the cavity (number of molecules in the polaritonic chemistry setup) is a generic point unrestricted to the Dicke model. As we have seen, in wide range couplings the photonic RPA conceptually provides a high quality solution of this problem. For the Dicke model, the corresponding xc functional is very simple and essentially analytical. However, fore realistic systems trying to implement and use the full OEP-RPA scheme of QEDFT does not look feasible, or at least feasibly. A possible compromise would be to combine the ideas of the first order photonic LDA of Ref.~\cite{Flick2022} with the ADCFD form of the photon RPA functional derived in this work.

\section*{Acknowledgement}
I.V.T. acknowledges support by Grupos Consolidados UPV/EHU del Gobierno Vasco (Grant No. IT1249-19) and by Spanish MICINN (Project No. PID2020-112811GB-I00).

\appendix
\begin{widetext}   
\allowdisplaybreaks
\section{Basic expressions and the lowest order diagrams for xc energy} 
In this section, we calculate several diagrams and diagrammatic elements entering the expressions for the xc energy in the main text. In particular, the KS polarizability 
$\Pi_s(\omega)$ diagrammatically can be written as:
\begin{equation} \label{Pi}
\Pi_s(i\omega)=\vcenter{\hbox{\begin{tikzpicture}[use Hobby shortcut, scale=0.6]
\draw (-0.8,0) .. (0.0,0.6) .. (0.8,0);
\draw (-0.8,0) .. (0.0,-0.6) .. (0.8,0);
\fill (-0.8,0) circle (1pt);
\fill (0.8,0) circle (1pt);
\end{tikzpicture}
}} \\
=\int\limits_{-\infty}^{\infty} \dfrac{d\epsilon}{2\pi}\sum\limits_{i,k=1}^{2}\dfrac{\lambda^2 d_{ki}d_{ik}}{(\textup{i}\epsilon - E_i)(\textup{i}(\epsilon+\omega) - E_k)}\\=\frac{4 T^2}{W \left(4W^2+\omega^2\right)}= -\dfrac{Z_s^{\textup{ex}}}{(i\omega)^2-\Omega_s^2},
\end{equation}
where energies $E_1$ and $E_2$ are equal to $\varepsilon_g$ and $\varepsilon_e$ respectively  (for the definition, see  Eq.~\eqref{eqn:KS_auxiliary}). 

The analytical expressions for the first order diagram $D_1$ and two second order diagrams $D_2$ and $D_3$ for the xc energy, shown in Fig.~\ref{fig:basic_diag} together with expansion coefficients, read  as follows,
\begin{align} \label{D1}
    D_1&=-\dfrac{1}{2}\dfrac{1}{(2\pi)^2}\int\limits_{-\infty}^{\infty}d\varepsilon_1 d\omega\sum\limits_{i=1}^2\sum\limits_{k=1}^2\dfrac{\lambda^2   d_{ik}d_{ki}}{(i\varepsilon_1-E_i)(i\varepsilon_1+i\omega-E_k)}D_{\textup{ph}}(i\omega)=\dfrac{\lambda^2 T^2}{W(\omega_{\textup{ph}}  + 2 W)},\\
    D_2&=-\dfrac{1}{4}\dfrac{1}{(2\pi)^3}\int\limits_{-\infty}^{\infty}d\varepsilon_1d\varepsilon_2 d\omega\sum\limits_{i=1}^2\sum\limits_{k=1}^2\sum\limits_{l=1}^2\sum\limits_{j=1}^2\dfrac{\lambda^4 d_{ik}d_{ki}d_{jl}d_{lj}}{(i\varepsilon_1-E_i)(i\varepsilon_1+i\omega'-E_k)(i\varepsilon_2-E_j)(i\varepsilon_2+i\omega'-E_l)}D_{\textup{ph}}^2(i\omega)\nonumber\\
    &=-\dfrac{\lambda^4 T^4}{W^2(\omega_{\textup{ph}} + 2 W)^3},\\
    \label{D3}
    D_3&=\dfrac{1}{4}\dfrac{1}{(2\pi)^3}\int\limits_{-\infty}^{\infty}d\varepsilon_1d\varepsilon_2 d\omega\sum\limits_{i=1}^2\sum\limits_{k=1}^2\sum\limits_{l=1}^2\sum\limits_{j=1}^2\dfrac{\lambda^4d_{ji}d_{ik}d_{kl}d_{lj}D_{\textup{ph}}(i\omega)D_{\textup{ph}}(i(\omega+\varepsilon_1-\varepsilon_2))}{(i\varepsilon_1-E_j)(i\varepsilon_1+i\omega-E_l)(i\varepsilon_2-i\omega-E_i)(i\varepsilon_2-E_k)}\nonumber\\
    &=\dfrac{\lambda^4 T^2 \left(\omega W+T^2\right)}{2W^2 (2 W+\omega_{\textup{ph}})^2(W+\omega_{\textup{ph}})},
\end{align}
where
\begin{align}
&d_{12} \equiv d_{ge}= d_{21} \equiv d_{eg}=\tripleship{\phi_g}{\hat{\sigma}_z}{\phi_e}=2vu=\dfrac{T}{W}=\sqrt{1-n^2},\\
&d_{11}\equiv d_{gg}=\tripleship{\phi_g}{\hat{\sigma}_z}{\phi_g}=-\dfrac{v_s}{W}=n,\quad d_{22}\equiv d_{e,e}=\tripleship{\phi_e}{\hat{\sigma}_z}{\phi_e}=\dfrac{v_s}{W}=-n.
\end{align}

\section{Explicit form of exchange energy in case  of two groups of equivalent  dimers}

\begin{multline}
E_\textup{xc}^{\rm RPA}\!=\!\int\limits_{0}^{\infty}\dfrac{d\omega}{2\pi}\!\textup{ln}\!\!\left[\!1\!-\!D_{\textup{ph}}(i\omega)\!\sum\limits_{j=1}^{2}\lambda_j^2  N_j\Pi_{sj}(i\omega)\!\right]\!\! d\omega\!=\!\dfrac{\left(\!\!-\omega_{\textup{ph}}\!-\!\Omega_{s1}\!-\!\Omega_{s2}\!+\!\sqrt{2 \sqrt{f_1 (e_3+f_2)-f_1^2}+(e_3+f_2)}\right)}{2},\nonumber
\end{multline}
where $f_1$, $f_2$, $e_3$ are  expressed as follows:
\begin{align} 
&f_1=\frac{3 e_2-e_3^2}{3 q_{e}}+\frac{e_3}{3}-\frac{q_{e}}{3}, \quad f_2=\frac{2 \sqrt{9 e_2^2-3 e_2 (2 e_3-q_{e}) (e_3+q_{e})+e_3^4+e_3^3 q_{e}+e_3 q_{e}^3+q_{e}^4}}{3 q_{e}},\nonumber\\
&e_1 = \omega_{\textup{ph}}^2 \Omega_{s1}^2 \Omega_{s2}^2,\quad e_2=Z_{s1}^{\textup{ex}} \Omega_{s2}^2+Z_{s2}^{\textup{ex}} \Omega_{s1}^2+\omega_{\textup{ph}}^2 \Omega_{s1}^2+\omega_{\textup{ph}}^2\Omega_{s2}^2+\Omega_{s1}^2 \Omega_{s2}^2,\quad e_3 = Z_{s1}^{\textup{ex}}+Z_{s2}^{\textup{ex}}+\omega_{\textup{ph}}^2+\Omega_{s1}^2+\Omega_{s2}^2,\nonumber\\
&q_{e}=\sqrt[3]{\dfrac{3 \sqrt{3} \sqrt{e_1 \left(27 e_1-9 e_2 e_3+2 e_3^3\right)-9 e_1 e_2 e_3+2 e_1 e_3^3+4 e_2^3-e_2^2 e_3^2}-\left(27 e_1-9 e_2 e_3+2 e_3^3\right)}{2}}.\nonumber
\end{align}
The asymptotic behavior of $E_{xc}^{\textup{RPA}}$ in $N_1$ and $N_2$ reads as: 
\begin{align}
E_\textup{xc}^{\rm RPA}\!&=\dfrac{\sqrt{Z_{s1}^{\textup{ex}}+Z_{s2}^{\textup{ex}}}}{2}+\frac{1}{2} \left(\sqrt{\dfrac{Z_1^{\textup{ex}} \Omega_{s2}^2+Z_2^{\textup{ex}} \Omega_{s1}^2}{Z_{s1}^{\textup{ex}}+Z_{s2}^{\textup{ex}}}}-\omega_{\textup{ph}}-\Omega_{s1}-\Omega_{s2}\right)\nonumber\\
&+\dfrac{(Z_{s1}^{\textup{ex}}+Z_{s2}^{\textup{ex}}) \left(2 \sqrt{\frac{\omega_{\textup{ph}}^2 \Omega_{s1}^2 \Omega_{s2}^2 (Z_{s1}^{\textup{ex}}+Z_{s2}^{\textup{ex}})}{Z_{s1}^{\textup{ex}} \Omega_{02}^2+Z_{s2}^{\textup{ex}} \Omega_{s1}^2}}+\omega_{\textup{ph}}^2+\Omega_{s1}^2+\Omega_{s2}^2\right)-Z_{s1}^{\textup{ex}} \Omega_{s2}^2-Z_{s2}^{\textup{ex}} \Omega_{s1}^2}{4 (Z_{s1}^{\textup{ex}}+Z_{s2}^{\textup{ex}})^{3/2}}+O\left(\dfrac{1}{n_1^l n_2^k}\right), \quad k+l=1.
\end{align}
Thus, the principal order is proportional to the root of the number of dimers. The asymptotics of interaction energy for an arbitrary number of groups can be determined by a simple formula:
\begin{align}
    E_{xc}^{\rm RPA}\!&=\dfrac{1}{2}\sqrt{\sum\limits_{j=1}^{N_{\textup{d}}}Z_{sj}^{\textup{ex}}}+O\left(1\right).
\end{align}

\end{widetext}

\bibliography{cavityQED.bib}

\end{document}